\documentclass[fleqn,usenatbib]{mnras}

\usepackage[T1]{fontenc}

\DeclareRobustCommand{\VAN}[3]{#2}
\let\VANthebibliography\thebibliography
\def\thebibliography{\DeclareRobustCommand{\VAN}[3]{##3}\VANthebibliography}

\usepackage{graphicx}	
\usepackage{amsmath}	
\usepackage{amssymb}	
\usepackage{nccmath}
\usepackage{epstopdf}
\usepackage{natbib}
\usepackage{subfig}
\usepackage{txfonts}
\usepackage{hyperref}
\usepackage[]{threeparttable}

\newcommand{\ve}[1]{{\rm\bf {#1}}}

\graphicspath{{Images/}}

\title[Reconnection driven cancellation and QSEB]{A reconnection driven magnetic flux cancellation and a quiet Sun Ellerman bomb}

\author[Kaithakkal et al.]{
Anjali. J. Kaithakkal,$^{1}$\thanks{E-mail: anjalijohnk@gmail.com}
J. M. Borrero,$^{2}$
A. Pastor Yabar,$^{3}$
and J. de la Cruz Rodr{\'i}guez$^{3}$
\\
$^{1}$Aryabhatta Research Institute of Observational Sciences, Nainital-263001, Uttarakhand, India \\
$^{2}$Leibniz Institute for Solar Physics, Sch\"oneckstrasse 6, 79104 Freiburg, Germany \\
$^{3}$ Institute for Solar Physics, Department of Astronomy, Stockholm University, AlbaNova University Centre, SE-106 91 Stockholm, Sweden
}

\date{Accepted 2023 March 09. Received 2023 February 14; in original form 2022 October 21}

\pubyear{2023}

\begin{document}
\label{firstpage}
\pagerange{\pageref{firstpage}--\pageref{lastpage}}
\maketitle

\begin{abstract}
The focus of this investigation is to quantify the conversion of magnetic to thermal energy initiated by a quiet Sun cancellation event and to explore the resulting dynamics from the interaction of the opposite polarity magnetic features. We used imaging spectroscopy in the H$\alpha$ line, along with spectropolarimetry in the \ion{Fe}{I} 6173~{\AA} and \ion{Ca}{II} 8542~{\AA} lines from the Swedish Solar Telescope (SST) to study a reconnection-related cancellation and the appearance of a quiet Sun Ellerman bomb (QSEB).  We observed, for the first time, QSEB signature in both the wings and core of the \ion{Fe}{I} 6173~{\AA} line. We also found that, at times, the \ion{Fe}{I} line-core intensity reaches higher values than the quiet Sun continuum intensity. From FIRTEZ-dz inversions of the Stokes profiles in \ion{Fe}{I} and \ion{Ca}{II} lines, we found enhanced temperature, with respect to the quiet Sun values, at the photospheric ($\log\tau_c$ = -1.5; $\sim$1000 K) and lower chromospheric heights ($\log\tau_c$ = -4.5; $\sim$360 K). From the calculation of total magnetic energy and thermal energy within these two layers it was confirmed that the magnetic energy released during the flux cancellation can support heating in the aforesaid height range. Further, the temperature stratification maps enabled us to identify cumulative effects of successive reconnection on temperature pattern, including recurring temperature enhancements. Similarly, Doppler velocity stratification maps revealed impacts on plasma flow pattern, such as a sudden change in the flow direction.
\end{abstract}




\section{Introduction}
Ellerman bombs (EBs) are transient line-wing brightenings with unaffected absorption at the line-core in the H$\alpha$ and other Balmer lines, which were first reported by \cite{1917ApJ....46..298E}. Since then their characteristics have been extensively investigated using observations \cite[see][for a review on EBs]{2013JPhCS.440a2007R}. EBs are generally observed in emerging flux regions or between evolving interspot regions \cite[][and references therein]{2002ApJ...575..506G,2008ApJ...684..736W,2016ApJ...823..110R}. They are reported to occur at magnetic flux cancellation sites \citep{2008PASJ...60..577M,2013A&A...557A.102B,2013ApJ...779..125N,2016ApJ...823..110R} and are considered to be driven by photospheric magnetic reconnection \citep{2011ApJ...736...71W,2013ApJ...774...32V,2019A&A...627A.101V}.\\

The first observational report on the quiet Sun counterpart of the EBs, the QSEBs, was presented by \cite{2016A&A...592A.100R}. QSEBs also, like EBs, are observed to be characteristic of photospheric reconnection \citep{2016A&A...592A.100R, 2018MNRAS.479.3274S,2020A&A...641L...5J}.  \cite{2018MNRAS.479.3274S} found that in addition to H$\alpha$ line-wing enhancement, QSEBs display a rise in the \ion{Ca}{II} 8542~{\AA} line-core and line-wing and a corresponding temperature increase in the upper photosphere. In contrast to EBs, QSEBs are shown to have lower intensity enhancement in the H$\alpha$ line-wing, $\sim$ 10 - 20 \%  with respect to the background intensity \citep{2018MNRAS.479.3274S}. QSEB signatures are also observed in the cores of \ion{Si} {IV} 1393~{\AA} and \ion{Si} {IV} 1403~{\AA}, and wings of \ion{C} {II} and \ion{Mg} {II} \citep{2017ApJ...845...16N}.\\

 Numerical modeling, using three-dimensional radiative magnetohydrodynamic simulations, was successful in recreating the observed spectral signatures of EBs \citep{2013ApJ...779..125N,2017ApJ...839...22H,2019A&A...626A..33H,2017A&A...601A.122D} and QSEBs \citep{2017A&A...601A.122D}, and in establishing that both phenomena are caused by photospheric magnetic reconnection. Using flux emergence simulations, \cite{2017ApJ...839...22H,2019A&A...626A..33H} showed that EBs are triggered by reconnection between $\cup$-shaped magnetic field lines. On the other hand \cite{2017A&A...601A.122D} showed EBs that form in $\Omega$-shaped field topology, using simulation of magnetic reconnection between an emerging magnetic field and the background magnetic field. \cite{2017A&A...601A.122D} also found that the flame-like morphology of EBs is visible only in limbward viewing and that in disk center observations they will be hard to be discerned from bright network magnetic features.\\

Magnetic field reconnection is at the heart of a majority of the dynamic and/or violent phenomena that we witness on the Sun. Magnetic flux cancellations, an observational signature of reconnection, occur in both  active and quiet regions of the Sun. In this work, we present such an event driven by a quiet Sun small-scale reconnection-related flux cancellation, namely the quiet Sun Ellerman bomb. We study in detail the temporal evolution of various quantities, like wing and core intensities, integrated Stokes $V$, and linear polarization signals, in the region where we observe the event. We then used spectropolarimetric inversion to quantify the associated physical parameters like vector magnetic field, line of sight velocity, and temperature to understand the magnetic field line topology, the plasma flow pattern, and the amount of local plasma heating driven by the event under consideration. We also estimated the total magnetic and thermal energy caused by the process to gain an understanding of the energy balance involved.

\section{Observations and analysis}
\label{sec:obsv}

A quiet Sun (QS) data set at the disk center was acquired on 2019 April 24, with the CRisp Imaging SpectroPolarimeter \citep{2008ApJ...689L..69S} at the 1-m Swedish Solar Telescope \citep[SST:][]{2003SPIE.4853..341S}. CRISP provided imaging spectropolarimetry \citep{2021AJ....161...89D} in the photospheric \ion{Fe}{I} 6173~{\AA} line (effective Land\'{e} factor $g_{\rm eff}$ = 2.5) at 15 spectral points between $\pm$~245~{m\AA} in step size of 35~{m\AA}. Co-spatial and nearly simultaneous spectropolarimetric data (i.e. all four Stokes parameters) in the \ion{Ca}{II} 8542~{\AA} line (effective Land\'{e} factor $g_{\rm eff}$ = 1.10) were recorded at 20 spectral line positions with varying step sizes: 75~{m\AA} close to the line-core, and 100~{m\AA} in the inner line-wings, plus a far wing point at $+$2.4~{\AA}. Simultaneous spectroscopic data (i.e. Stokes $I$ only) in the Balmer H$\alpha$ 6563~{\AA} line at 25 spectral points between $\pm$~1.2~{\AA} with a step size of 100~{m\AA} were also recorded for the same region. The data set has a mean cadence of 28.2~s.\\

\begin{figure*}
\centering
  \includegraphics[trim=0 60 15 65,clip,width=1.0\textwidth]{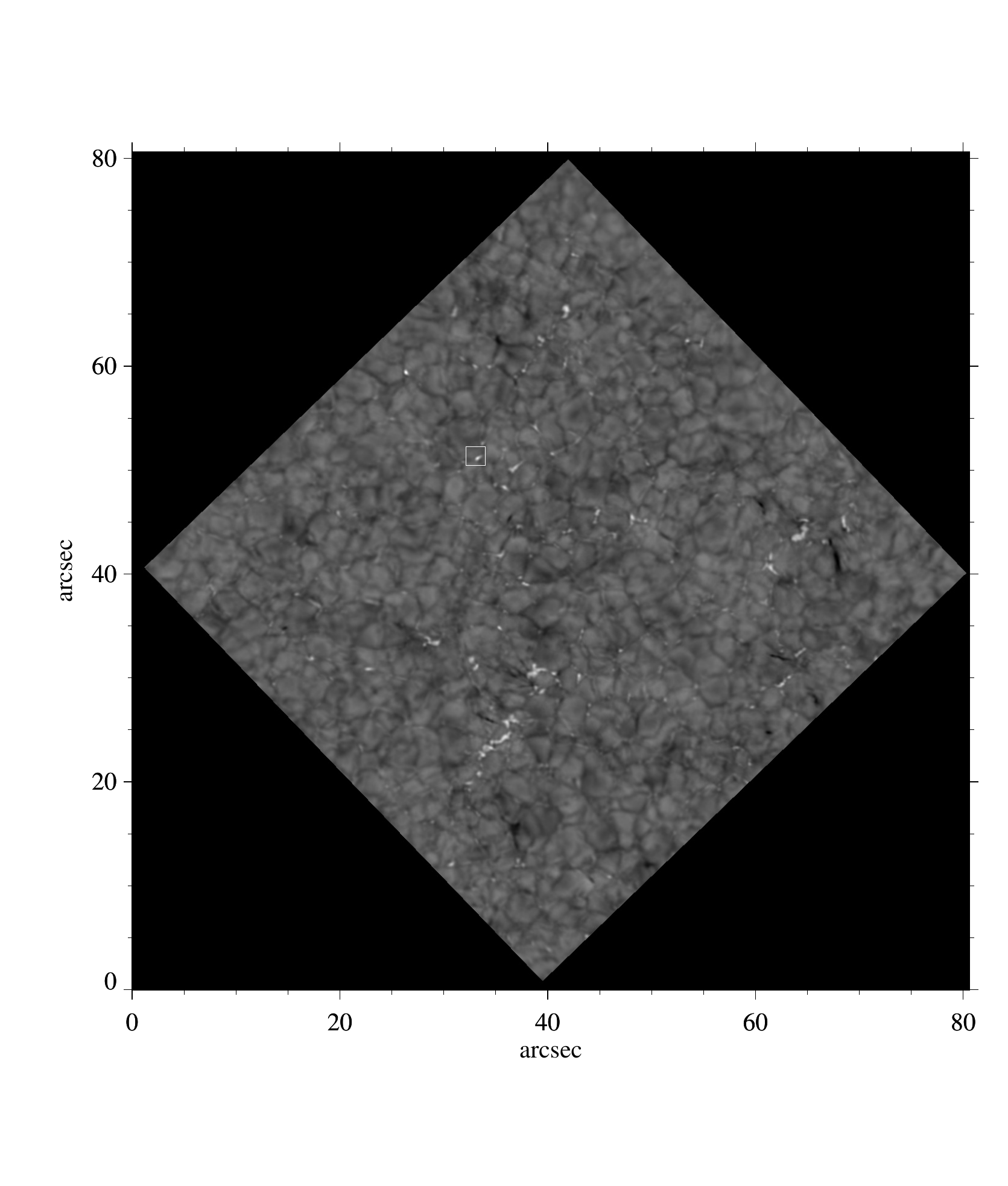}
  \caption{Full FOV at $-$1.2~{\AA} from the H${\alpha}$ line core. The white box indicates our 5.95"$\times$5.95" region of interest. The frame is taken at 9:49:51 UT.}
  \label{fig:a}
\end{figure*}

CRISP has  an image scale of 0.059\arcsec per pixel and covers a field of view (FOV) of  $\sim$ 55\arcsec $\times$ 55\arcsec (see Fig.~\ref{fig:a}). The data calibration was carried out using the SSTRED pipeline \citep{2015A&A...573A..40D,2021A&A...653A..68L}, and was further processed with the multi-object multi-frame blind deconvolution code \citep[MOMFBD,][]{2005SoPh..228..191V}.\\

With the help of the Fourier Transform Spectrometer \citep[FTS; ][]{1999SoPh..184..421N} atlas in the \ion{Ca}{II} region we found that the wing intensity recorded at $+$2.4~{\AA} from the line core is only 76\% of the real continuum. We did the same analysis for the H$\alpha$ line and obtained that the wing intensity at $\pm~$1.2~{\AA} from the line core is only 81\% of the continuum. All these corrections are applied to the Stokes profiles of respective spectral lines. The Stokes profiles are then normalized to the spatiotemporal average QS intensity, $<I_\mathrm{QS}>$, of respective spectral lines.\\

The noise level ($\sigma$) in Stokes $Q$, $U$, and $V$, determined using the polarization signals at the farthest wavelength points from line core, are 1.9 $\times$10$^{-3}$, 1.6 $\times$ 10$^{-3}$, and 1.7 $\times$ 10$^{-3}$ respectively, for the \ion{Ca}{II} 854.2~{nm} line and 1.4 $\times$ 10$^{-3}$ for the \ion{Fe}{I} 617.3~{nm} line, in units of $<I_\mathrm{QS}>$.\\

\subsection{Region of Interest (ROI)}
\label{subsec:roi}

We chose an ROI of size 5.95\arcsec$\times$ 5.95\arcsec (white box in Fig.~\ref{fig:a}) which covered a small-scale magnetic flux cancellation event. The event apparently lasted for about 39.0 minutes, by the end of which the smaller magnetic feature, which is of negative polarity, went below our threshold of 3 $\sigma$ value of Stokes $V$ noise. We identified the opposite polarity features and tracked them, from the \ion{Fe}{I} 6173~{\AA} line Stokes $V$ maps, based on spatial overlap \citep{2019A&A...622A.200K}. Our code first identified the negative polarity patch, spatially close to the positive patch, 17 minutes after the positive polarity patch was detected. So this cancellation event appears to be resulting from the interaction between a relatively big existing positive-polarity feature and a rather small emerging negative-polarity feature.\\

A number of ancillary quantities, that will be very useful later, were derived from the observed Stokes vector. These are (a) continuum maps of \ion{Fe}{I} 6173~{\AA} line constructed by averaging Stokes $I$ values at $\pm$~0.245~{\AA}; (b) H$\alpha$ wing intensity maps made from the mean of intensities at $\pm$~1.2~{\AA} from the line core; (c) the wavelength-integrated absolute value of the circular polarization $V$ as well as wavelength-integrated linear polarization (LP =$\sqrt{Q^2+U^2}$) maps for the \ion{Fe}{I} 6173~{\AA} and \ion{Ca}{II} 854.2~{nm} line.\\

\subsection{FIRTEZ-dz Inversion} 
\label{sec:inv}

In order to extract the physical parameters of the solar atmosphere from the observations of the Stokes vector described above, we employ the FIRTEZ-dz inversion code \citep{adur2019firtez}. FIRTEZ-dz calculates the gas pressure $P_{\rm g}$ consistently with the magnetohydrostatic equations including the effects of the Lorentz force \citep{borrero2021mhs}. This is unlike inversions codes such as SIR \citep{basilio1992sir}, SPINOR \citep{frutiger1999spinor}, NICOLE \citep{hector2015nicole}, SNAPI \citep{milic2018} and STiC \citep{jaime2019stic}, where the gas pressure $P_{\rm g}$ is obtained under the assumption of hydrostatic equilibrium. Because of this, FIRTEZ-dz is capable of working directly in the geometrical scale $z$ instead of the continuum optical-depth $\tau_c$.\\

In our inversions in this work, the initial atmosphere is thnone VALC model \citep{vernazza1981}, from where we take $T(z)$, $P_{\rm g}(z)$, and $\rho(z)$ between $z=[0 - 1536$]~km interpolated to an equidistant $z$-grid with $\Delta z=12$~km and expanded by assuming a plane-parallel atmosphere into the entire $(x,y,z)$ domain. Since VALC does not include any values for either the magnetic field vector $\ve{B}(z)$ or the line-of-sight velocity $v_{\rm los}(z)$, these parameters are initially estimated at each $(x,y)$ pixel by applying the weak-field approximation to the observed Stokes parameters in the Fe I 6173~{\AA} line \citep{jefferies1991} and by calculating the center-of-gravity of Stokes $I$ in the same spectral line. The values thus obtained for $\ve{B}$ and $v_{\rm los}$ are initially considered to be constant with $z$.\\

The modifications to the physical parameters at each iteration step of the inversion process at the selected $z$ locations (i.e. nodes) are as follows: 64 for $T(z)$, 32 for $v_{\rm los}(z)$, and 2 for each of the three components of the magnetic field vector $B_{\rm x}(z)$, $B_{\rm y}(z)$, and $B_{\rm z}(z)$. Each node is considered a free parameter during the inversion process.  The large number of free parameters in the temperature is needed so as to ensure that the geometrical distance between the $\log\tau_c=0$ level and the nearest node is not too large, otherwise the inversion finds it difficult to correctly reproduce the observed continuum intensity in the observations. In order to avoid lack of convergence as a consequence of the large number of free parameters employed, we use a Tikhonov regularization \citep{tikhonov1995} that penalizes those solutions where the vertical derivatives of the physical parameters are too large \citep{jaime2019stic}. During the inversion different weights $w$ are given to each of the four components of the Stokes vector: $\mathbb{I}=(I, Q, U, V)$. In our case we used: $w_1=1$, $w_2=4$, $w_3=4$, $w_4=4$, for Stokes $I$, $Q$, $U$, and $V$, respectively.\\

FIRTEZ-dz typically operates under the assumption of Local Thermodynamic Equilibrium (LTE). This is indeed a reasonable assumption for the \ion{Fe}{I} 6173~{\AA} line. However, in order to improve the inference of the temperature stratification in the upper photosphere and lower chromosphere, it is required to consider non-LTE effects in the treatment of the \ion{Ca}{II} 8542~{\AA} line \citep[see, e.g.][]{2000ApJ...530..977S,jaime2012invnlte}.  Therefore, in this paper, the FIRTEZ-dz inversion code has been modified to account for deviations from Local Thermodynamic Equilibrium by implementing departure coefficients for the upper, $\beta_{\rm u}$, and lower levels, $\beta_{\rm l}$, involved in the electron transition that gives rise to a particular spectral line (i.e. \ion{Ca}{II} 8542~{\AA}). To this end, we follow the formulation by \cite{hector1998} to determine the line-opacity and source function, as well as their derivatives with respect to physical parameters, under the so-called \emph{Fixed departure coefficients} (FDC) approximation whereby the departure coefficients are not modified during the inversion even though the physical parameters (mainly the temperature) are being modified. Under this approximation, the departure coefficients and therefore the level populations under non-LTE are determined only once at the beginning of the inversion process described above.  This is therefore a first order non-LTE rather than a full non-LTE inversion.\\

The departure coefficients are determined for the VALC model by solving the Statistical Equilibrium equations with a \ion{Ca}{II} atomic model that includes five bound states and one continuum level. The Gotrian diagram of the atomic model can be found in \citet[][see Fig.~1]{shine1974}. We note that the equation-of-state that is used to determine the electron pressure is still solved under LTE. The angular integration that is needed to calculate the mean intensity of the radiation field is performed by assuming that the 1D model is plane-parallel and uses 10 ray paths with five angles and two directions (incoming and outgoing) for each angle.\\

\begin{figure*}[h]
\centering
  \includegraphics[width=0.9\textwidth]{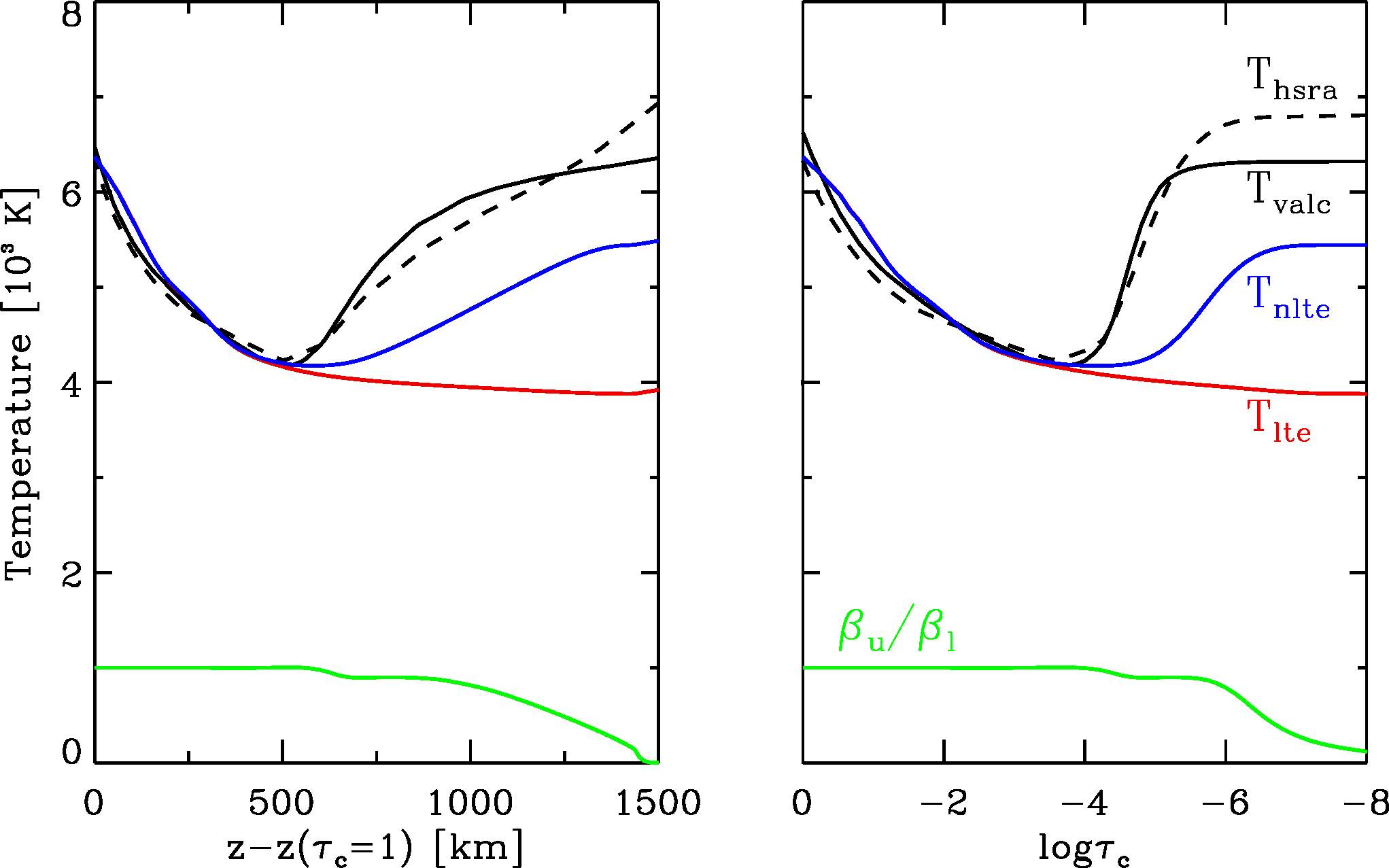} \\
  \includegraphics[width=0.9\textwidth]{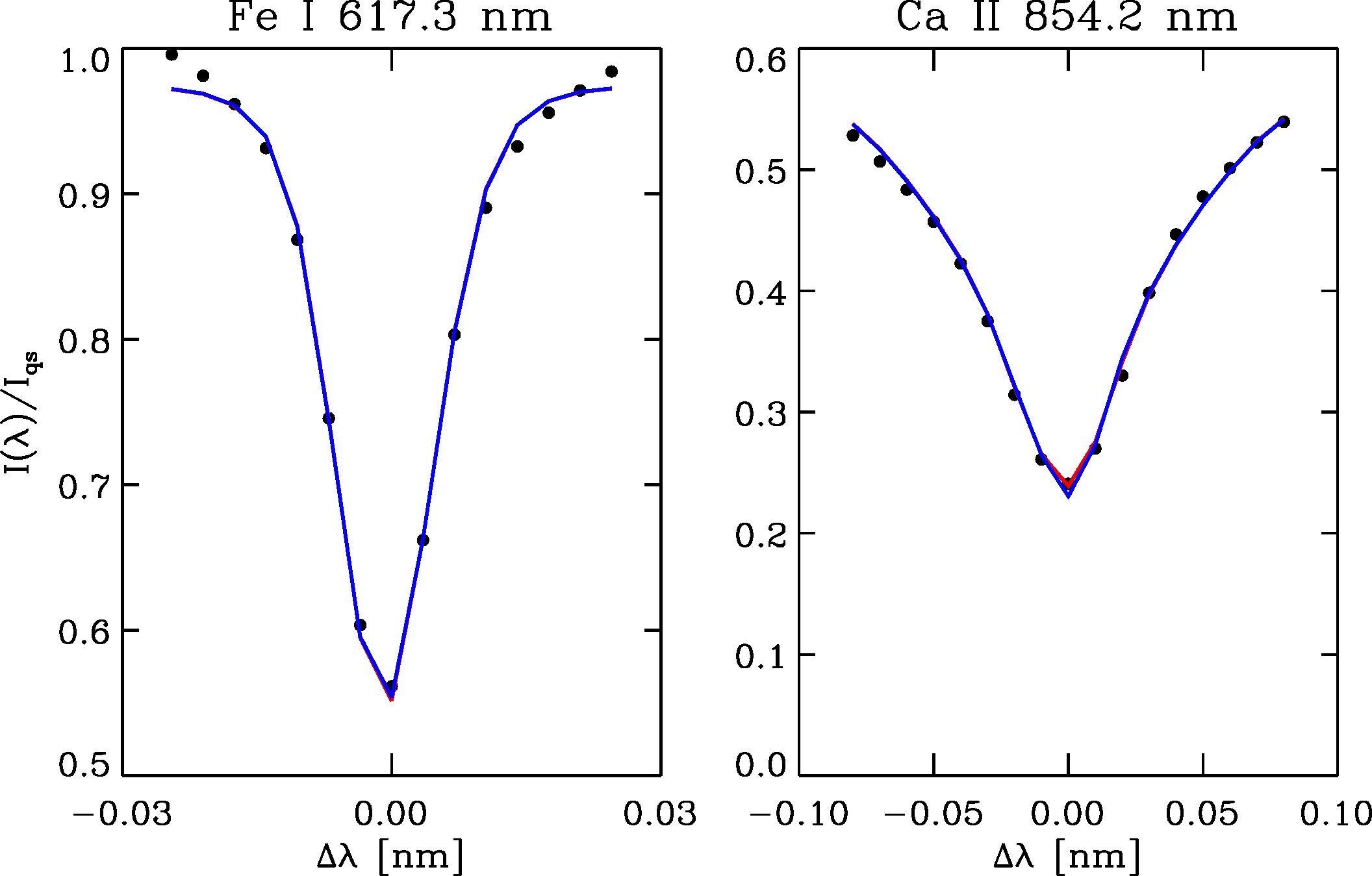}
  \caption{{\it Upper panels}: average temperature stratification as a function of geometrical height $z$ (left) and optical-depth $\tau_c$ (right) obtained from the inversion of all individual Stokes profiles in the selected ROI (Sect.~\ref{subsec:roi}). Red and blue curves correspond to the LTE and non-LTE inversions respectively. Solid-black and Dashed-black lines display the temperature stratifications from the VALC and HSRA models respectively. {\it Bottom panels}: average observed intensity profiles ($I(\lambda)$; filled circles) and best-fit profiles produced by the LTE (red) and non-LTE (blue) inversions for the \ion{Fe}{I} (left) and \ion{Ca}{II} lines.}
  \label{fig:lte_vs_nlte_nohydro}
\end{figure*}

This inversion code is applied to the observations of the \ion{Fe}{I} 6173~{\AA} and \ion{Ca}{II} 8542~{\AA} only. The \ion{H}{I}~6563~{\AA} is neglected at this point for two main reasons. The first one is that for this spectral line we do not have polarization signals (only Stokes $I$), and the second one is that FIRTEZ-dz considers only 1D radiative transfer while the aforementioned spectral line is heavily affected by 3D radiative transfer effects \citep{leernaarts2012}.\\

The inversion is performed considering the transmission profiles for the CRISP instrument for each spectral line, \ion{Fe}{I} and \ion{Ca}{II}, independently. The transmission profiles are theoretically calculated by considering the individual contribution of each etalon from the measured wavelength-dependent reflectivity and distance between plates \citep{2006A&A...447.1111S,2015A&A...573A..40D}. In addition, before the inversion, the Stokes vector in the \ion{Ca}{II} 8542~{\AA} line is re-sampled from the original grid described above to a 100~m{\AA} evenly-spaced wavelength grid from $-800$~m{\AA} to $+800$~m{\AA} with a total of 17 wavelength points. The \ion{Fe}{1} 6173~{\AA} spectral line is not re-sampled because it was already recorded with an constant step size. The atomic data for the spectral lines inverted by FIRTEZ-dz are presented in Table~\ref{tab:atomic_data}.\\

\begin{table*}
\begin{threeparttable}
\caption{Spectral lines and their associated atomic parameters used in the inversion.}
\label{tab:atomic_data}
\centering                 
\begin{tabular}{cccccccccc}
\hline
\hline                     
$\lambda_0$ & $\Delta\lambda$ & ${\rm n}_{\lambda}$ & ${\rm e}_{\rm low}$ & ${\rm e}_{\rm upp}$ & 
$\log_{10}{{\rm g}\,{\rm f}}$ & ${\rm E}_{\rm low}$ & $\alpha$ & $\sigma/{\rm a}_{0}^{2}$ \\
 $\textrm{[nm]}$ & [m{\AA}] &  & &  &  &  [eV] & &  &  \\
\hline                     
    Fe {\sc I} 617.33 & 35 & 15 & $^{5.0}{\rm P}_{1.0}$ & $^{5.0}{\rm D}_{0.0}$ & -2.880 & 2.223 & 0.266 & 280.57\\
    Ca {\sc II} 854.21 & 100 & 17 & $^{2.0}{\rm D}_{2.5}$ & $^{2.0}{\rm P}_{1.5}$ & -0.360 & 0.000 & 0.275 & 291.00\\
\hline                                           
\end{tabular}
\begin{tablenotes}
      \small
      \item $\lambda_{0}$ is the central wavelength for the electronic transition associated to the spectral line. $\Delta\lambda$ is the pixel size in m{\AA}, ${\rm n}_{\lambda}$ is the number of wavelengths used for that spectral line. ${\rm e}_{\rm low}$ and ${\rm e}_{\rm upp}$ are the electronic configurations of the lower and upper energy levels, respectively. ${\rm E}_{\rm low}$ is the excitation potential (in eV) of the lower energy level. The atomic data for \ion{Fe}{I} have been adopted from \cite{nave1994}, whereas for \ion{Ca}{II} we employ \citep{edlen1956} (see also Atomic Spectra Database - ASD - of the National Institute of Standards and Technology - NIST). $\alpha$ and $\sigma/{\rm a}_{0}^{2}$ are the velocity exponent and collision cross-section parameters, respectively as they are defined in Anstee, Barklem and O'Mara collision theory for the broadening of metallic lines by neutral hydrogen collisions \citep{anstee1995, barklem1997, barklem1998}.
\end{tablenotes}
\end{threeparttable}
\end{table*}

\subsection{Comments on the 180$^{\circ}$ resolution}
\label{subsec:180amb}

Although it has not been mentioned above, \citet{borrero2021mhs} (see their Fig.~2) indicates that before the gas pressure is computed via the application of MHS equilibrium, it is necessary to correct the 180$^{\circ}$-ambiguity that affects the $B_{\rm x}(x,y,z)$ and $B_{\rm y}(x,y,z)$ components of the magnetic field inferred by FIRTEZ-dz. This is typically done via the non-potential field calculation \citep[NPFC;][]{manolis2005}. Unfortunately, the quiet Sun is characterized by low signal-to-noise ratios in the linear polarization (i.e. Stokes $Q$ and $U$; see Sect.~\ref{sec:obsv}), thereby hindering the 180$^{\circ}$-ambiguity correction. This is clear from the moment we study the time consistency in the corrected $B_{\rm x}$ and $B_{\rm y}$. The NPFC method yields disambiguation such that the horizontal field at a given location oftentimes changes direction between consecutive time frames. To address this issue we select $\ve{B}(x,y,z,t=$~9:46:58~UT), where the NPFC yields a spatially consistent result. From there we disambiguate, using a simple acute-angle method \citep{cheung2008}, frame at $t-\Delta t$. Once $t-\Delta t$ is corrected, we use this to correct the 180$^{\circ}$-ambiguity at $t-2\Delta t$. We proceed in this fashion until we reach $t=0$. We then restart again from $\ve{B}(x,y,z,t=$~9:46:58~UT) and move forward in time: $t+\Delta t$, $t+2\Delta t$, and so forth.\\

\subsection{Average quiet Sun temperature: LTE vs. non-LTE inversion}
\label{subsec:average_qs}

In \citet{2020A&A...634A.131K} we attempted to assess the temperature enhancement during a reconnection event, in the upper photosphere (i.e. temperature minimum) close to $\log\tau_c \approx -3$ from the inversion, under Local Thermodynamic Equilibrium (LTE), of the \ion{Si}{I} 10827~{\AA} spectral line. However, this spectral line already shows important deviations from LTE at this height \citep[see Fig.~6 in][]{natalia2018} and therefore the temperature estimations were only approximate. As already mentioned, in this paper we have decided to include a first-order non-LTE treatment by including fixed departure coefficients for the \ion{Ca}{II} 8542~{\AA} line. To demonstrate the capability of FIRTEZ-dz in retrieving more accurate temperatures in the upper photosphere and lower chromosphere, we show in Figure~\ref{fig:lte_vs_nlte_nohydro} the average temperature, as a function of geometrical height $z$ (upper-left) and optical depth (upper-right), obtained from the inversion of all spatial pixels in the region of interest (Sect.~\ref{subsec:roi}) at all observed times. Therefore this average was constructed from the average of more than 1 million Stokes $ I(\lambda)$ profiles. The results from this approximate non-LTE inversion are shown in blue, whereas results under LTE are shown in red. For comparison purposes, we show also $T(z)$ from two semi-empirical models: HSRA \citep[dashed black line;][]{gingerich1971}, and VALC \citep[solid black line;][]{vernazza1981}. On the one hand, the temperature retrieved from the LTE inversion shows a monotonic decrease without either temperature minimum or chromospheric temperature increase. Our non-LTE inversion, on the other hand, features both temperature minimum and chromospheric increase for $z-z(\tau_c=1) > 500$~km or $\log\tau_c < -4$. The reason the inversion infers a temperature increase in the chromosphere is that the ratio between the upper and lower departure coefficients in the \ion{Ca}{II} 8542~{\AA} line drops below 1: $\beta_{\rm u}/\beta_{\rm l} < 1$ (see green lines in the upper panels of Fig.~\ref{fig:lte_vs_nlte_nohydro}). Since the source function is roughly proportional to $S \propto \beta_{\rm u}/\beta_{\rm l} \exp[{-hc/2KT}]$, as the ratio $\beta_{\rm u}/\beta_{\rm l}$ drops, the temperature must increase so as to yield the same amount of emitted energy per unit time ($s$), per unit area (m$^2$), and per solid-angle unit (steradian) as in the LTE case.\\

We note that, despite having very different temperature stratifications, $T(z)$ or $T(\tau_c)$, both LTE (red) and non-LTE (blue) inversions are equally capable of fitting the observed (spatially averaged) intensity profiles of both spectral lines in Table~\ref{tab:atomic_data}.\\

Although the chromospheric temperature from the approximate non-LTE inversion is much closer to that of the semi-empirical models than that from the LTE inversion, it is still somewhat lower. There are several possible reasons for this. One reason is that, although the source function is modified to include non-LTE effects, the Saha equation (i.e. LTE) is still employed to determine the electron density (Sect.~\ref{sec:inv}). The Saha equation tends to overestimate, compared to non-LTE, the electron density and therefore LTE needs lower temperatures to yield the same ionization level \citep{silva2018,2019A&A...627A.101V}.  Another reason is of course the fact that we are employing fixed departure coefficients, namely those obtained from the VALC model, in our inversions. This approach limits the accuracy of our inferred temperatures. We estimated the error in the temperature, introduced by our approximate first-order non-LTE treatment, by restarting the inversion process with updated departure coefficients, obtained from the final models after the first inversion, on a small $10\times 10$ pixels region over a limited number of timesteps. We then compared the resulting temperatures with those obtained with the fixed departure coefficients from the VALC model. The results indicate that the temperature remains the same up to $\log\tau_c=-3.5$, but that fixing the departure coefficients to those from VALC systematically underestimates the temperatures above this layer. The underestimation is of about 100~K at $\log\tau_c=-4$ and quickly raises towards higher layers. At the highest point considered in our analysis, $\log\tau_c=-4.5$ (see Sect.~\ref{sec:invr}), the underestimation is already 500~K.\\


\section{Results} 
\label{sec:re}
\subsection{Qualitative Description}
\label{sec:re1}

\begin{figure*}
\centering
  \includegraphics[trim=1 40 5 30,clip,width=1.0\textwidth]{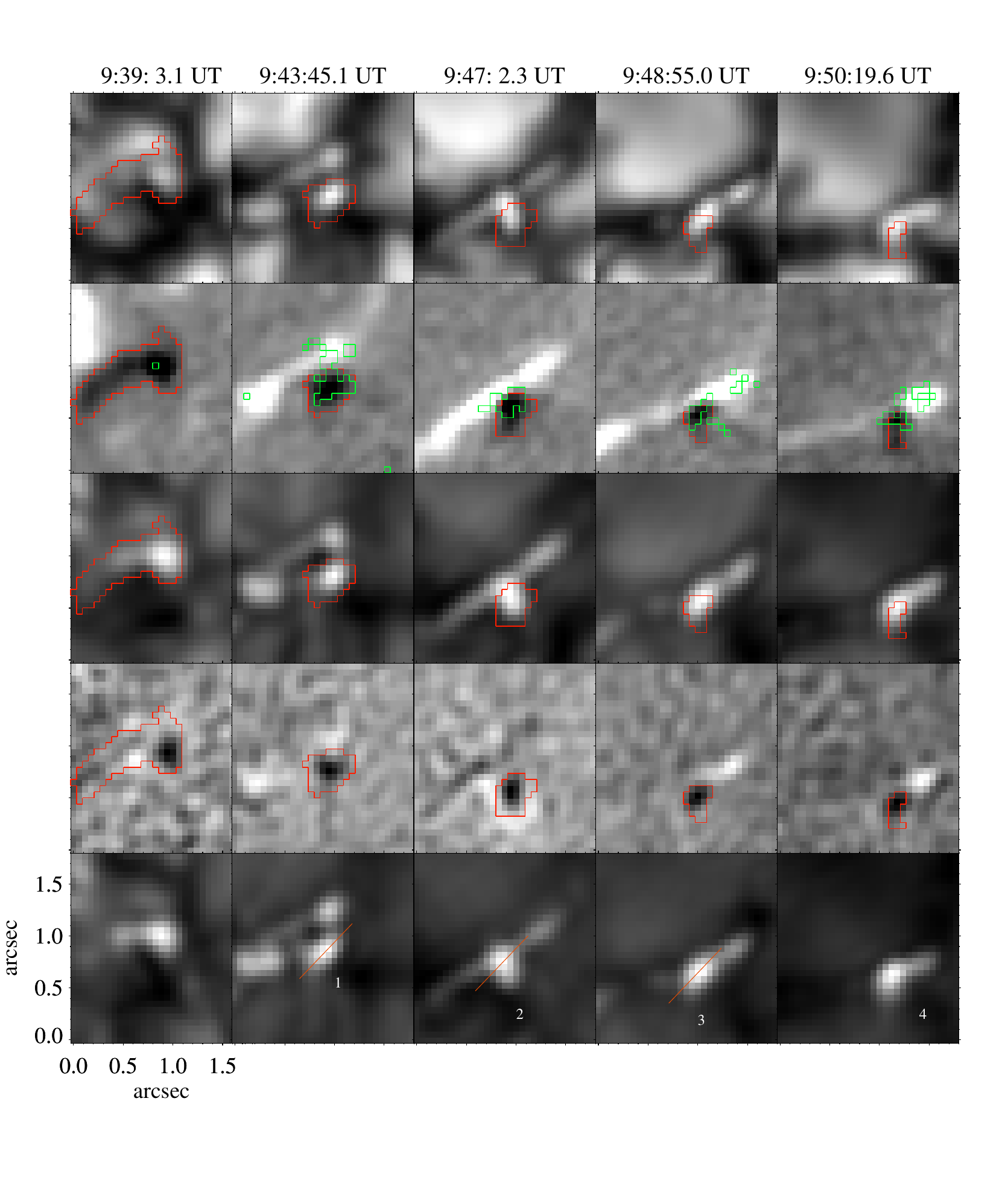}
  \caption{{\it From top to bottom}: continuum intensity in \ion{Fe}{I} 6173~{\AA} (first row), Stokes $V$ in \ion{Fe}{I} 6173~{\AA} with the green contours outlining regions with linear polarization signal greater than 3$\sigma$ value in this line (second row), wing intensity in \ion{Ca}{II} 8542~{\AA} (third row), Stokes $V$ in \ion{Ca}{II} 8542~{\AA} (fourth row), and finally, blue wing intensity in H$\alpha$ at -1.2~{\AA} from the line core (fifth row). The red contour outlines the negative polarity patch which is skipped in the last row to provide clear visibility of the QSEB's presence. The red cuts in the H$\alpha$ maps where brightening enhancements denoted as 1,2, and 3 are used to make maps given in Figs.\ref{fig:i}, \ref{fig:j}, and \ref{fig:k}. The direction of such cuts is referred to as $L$. A movie of the event evolution in the continuum and Stokes $V$ of the \ion{Fe}{I} line and in the blue wing intensity of H$\alpha$ line at -1.2~{\AA} is available online.}
  \label{fig:b}
\end{figure*}

Out of the 39.0 minutes-long apparent lifetime of the negative polarity feature involved in the cancellation event, we focus on the last 13.15 minutes of evolution in the present work.  The presence of a QSEB during the event is evident from our observation. We note that the H$\alpha$ spectral line exhibits enhanced inner and outer wings with an unaffected line core (see right column of Fig.~\ref{fig:d}) and that the wing intensities rise recurrently during cancellation (see last row of Fig.~\ref{fig:b}). The characteristic flame-like morphology of the QSEB is not visible in our data as the observation is taken at disk center \citep[see][for details]{2017A&A...601A.122D}.\\

The evolution of the event is demonstrated in Fig.~\ref{fig:b} using continuum/wing intensity and wavelength-integrated Stokes $V$ maps, created from the observed spectral lines, at different time steps. From the \ion{Fe}{I} continuum intensity map (first row) we see that the intensity within the negative patch rises before finally falling off towards the end. The Stokes $V$ map (second row) clearly shows that the negative patch is interacting with a relatively bigger positive patch that splits and merges during the evolution, and the negative patch undergoes a steady and continuous shrinkage with time.  We also observe significant linear polarization in the \ion{Fe}{I} line between the interacting magnetic features, which indicates their magnetic connectivity. From the third row, it is clear that the \ion{Ca}{II} wing intensity at +2.4~{\AA} from the line core is also rising, and falling during the event. Stokes $V$ maps in the \ion{Ca}{II} line (fourth row) display clear signals above the 3$\sigma$ level. However, these maps look rather diffused and the evolution is uneventful. Maps of the H$\alpha$ wing intensity at -1.2~{\AA} (i.e. blue-wing) from line-core (fifth row) show the presence of enhanced brightness. We denote some of these regions of enhanced brightness at different time steps as 1, 2, 3, and 4. While 1 is fully covered by the negative patch contour, 2, 3, and 4 fall mostly into the positive polarity patch region. The brightening enhancement is the highest for the 4th one (t~=~9:50:19.6 UT) with the H$\alpha$ blue wing intensity reaching values almost 40 \% higher than the corresponding mean QS value. The other three regions feature somewhat smaller intensity enhancements: about 30-35 \% larger than the average QS signal.\\

\begin{figure*}
  $\begin{array}{rl}
  {\includegraphics[trim=20 115 110 70,clip,width=0.5\textwidth]{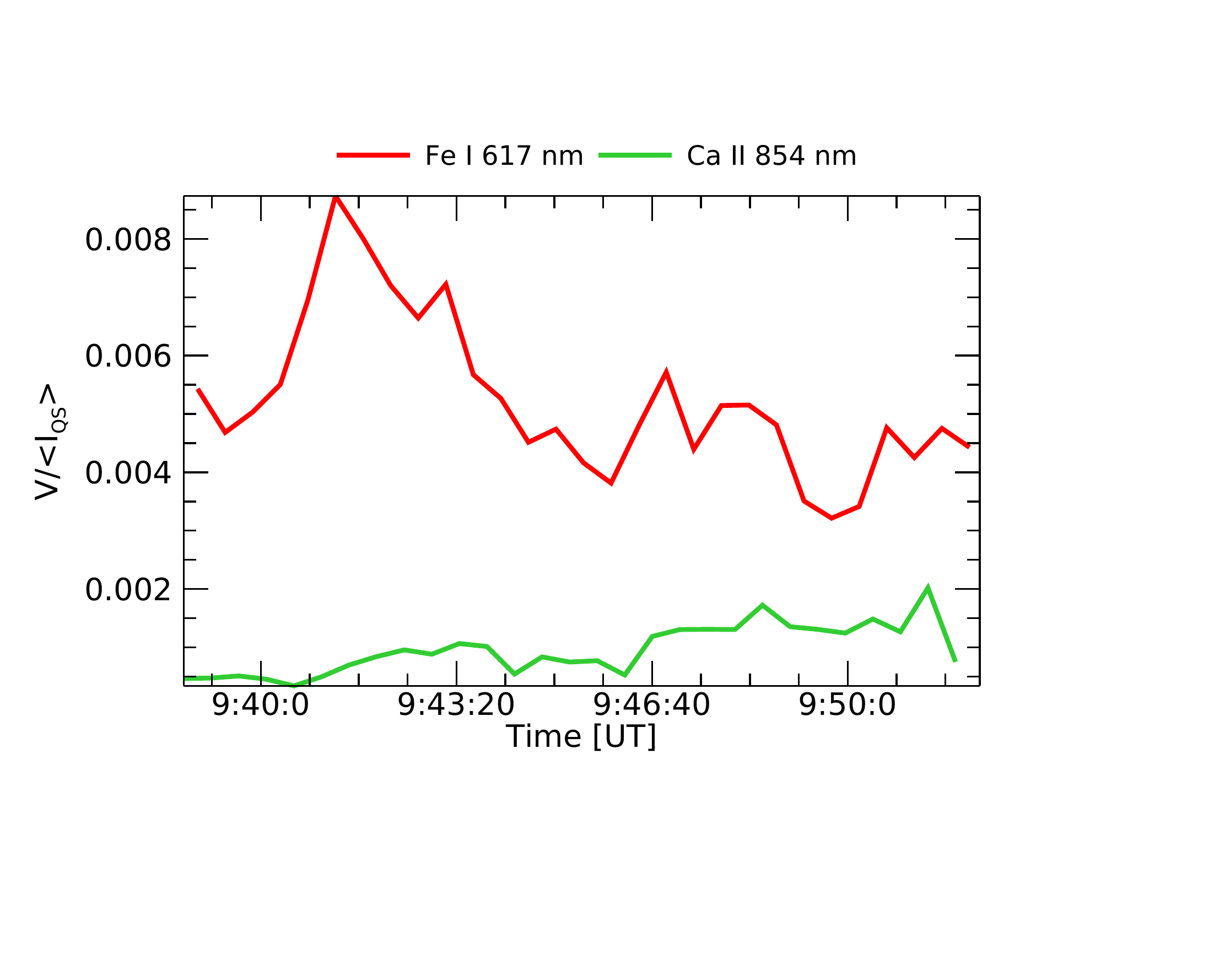}}&
   {\includegraphics[trim=20 115 110 70,clip,width=0.5\textwidth]{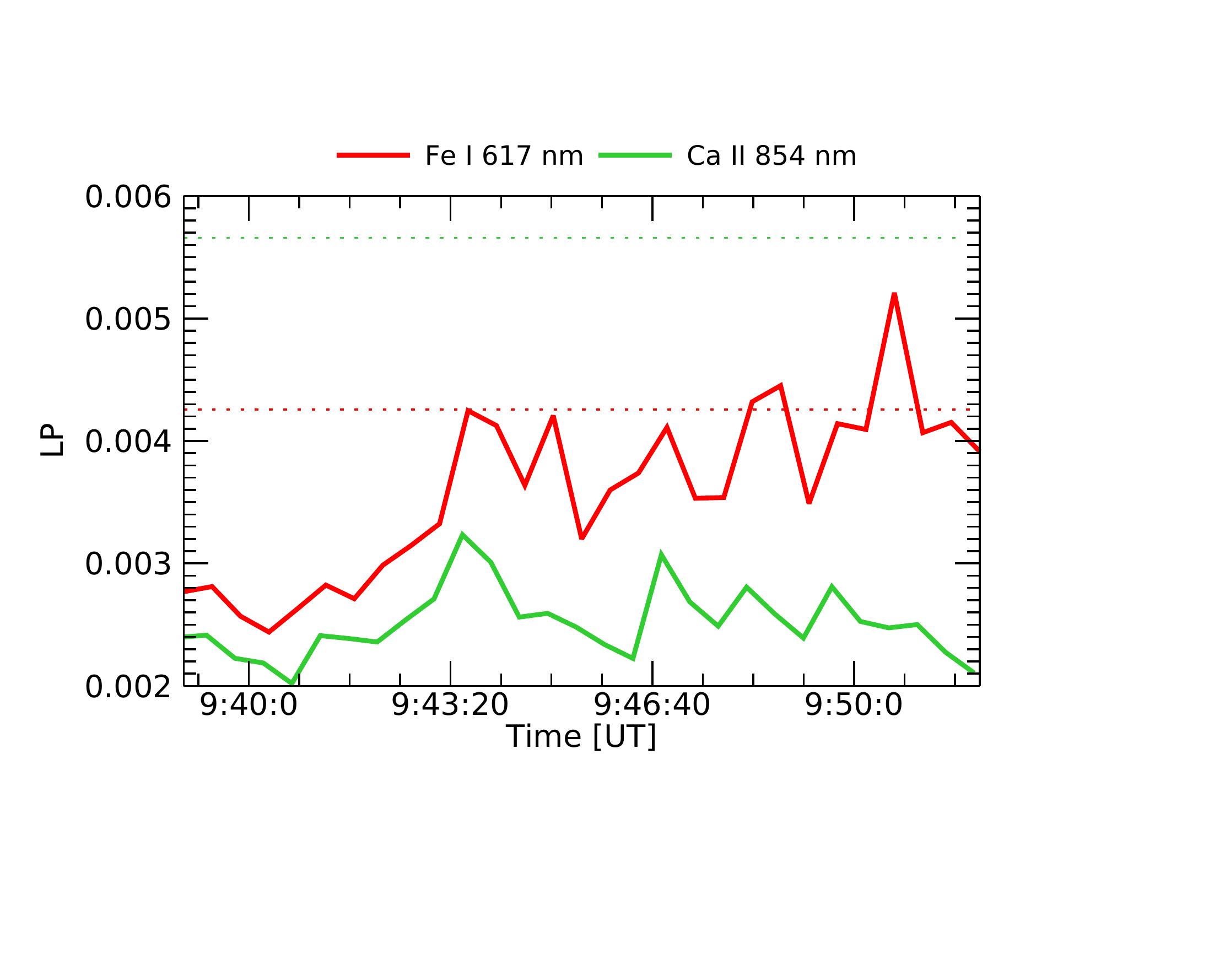}}
 \end{array}$
  \caption{Plots in red and green represent curves for \ion{Fe}{I} and \ion{Ca}{II}, respectively. Left: evolution of Stokes $V$; Right: linear polarization (LP) - the dotted lines represent respective 3$\sigma$ values.
 \label{fig:c}}
\end{figure*}

\begin{figure*}
\centering
  \includegraphics[trim=1 85 80 1,clip,width=1.0\textwidth]{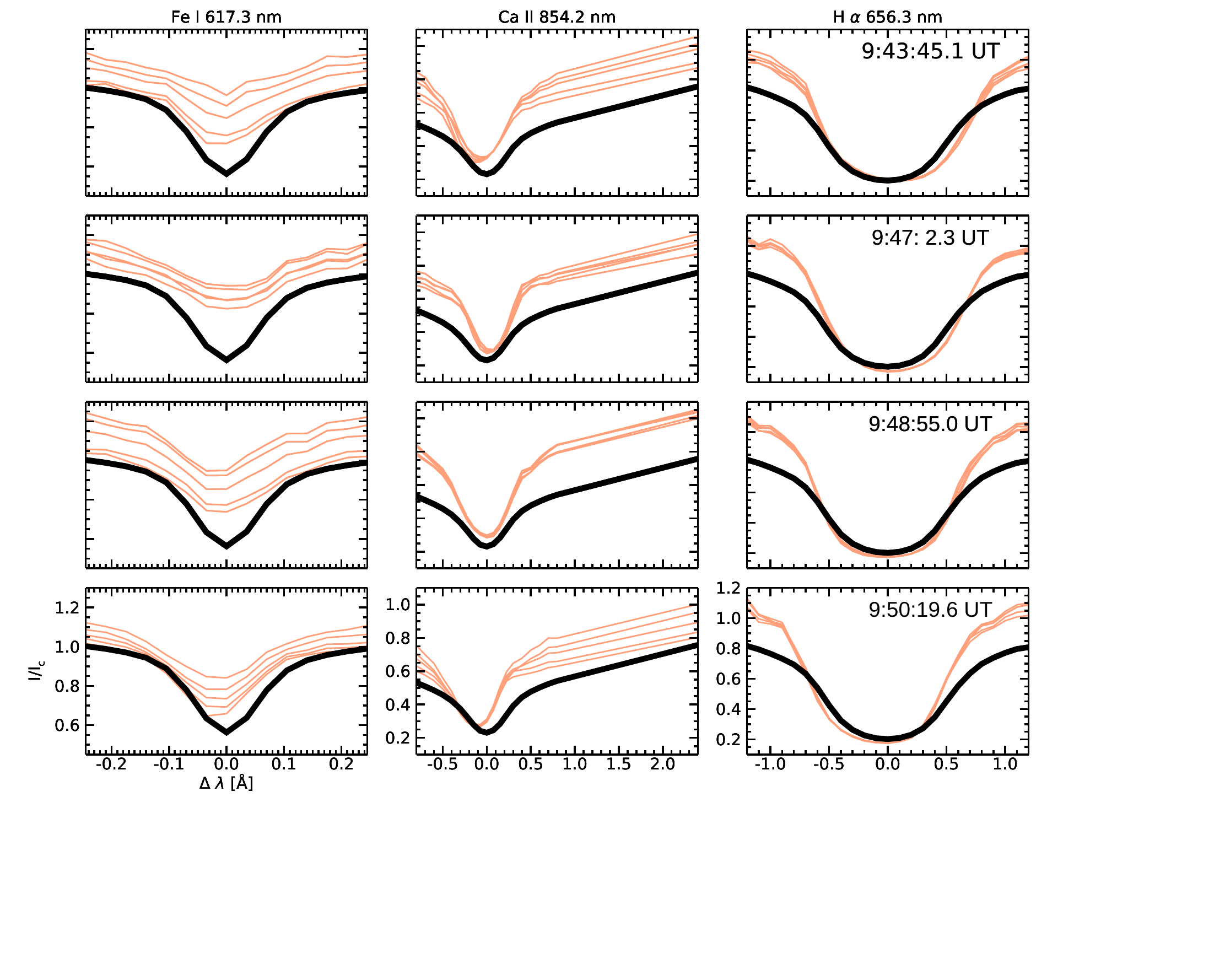}
  \caption{{\it From left to right}: Stokes $I$ profiles of \ion{Fe}{I}, \ion{Ca}{II} and H$\alpha$ at different time steps corresponding to the numbered brightenings in Fig.~\ref{fig:b}. For all the plots, the black curves represent the spatiotemporal mean QS profile for the respective spectral lines. And the orange curves are the profiles from the five-pixel region (see text for details) at the corresponding time step. The \ion{Ca}{II} profiles are plotted in their original observed spectral range. The timestamp in the rightmost plot is the time at which the blue wing of H$\alpha$ is recorded.}
  \label{fig:d}
\end{figure*}

Fig.~\ref{fig:c} presents the temporal evolution of Stokes $V$ and LP signals of the negative patch.  We chose the negative polarity patch as the representative feature as it stayed compact (i.e. it did not break into several patches or interact with other like-polarity features) during the cancellation event, thus making it easier to track its evolution. We took the mean of the wavelength integrated Stokes $V$ signal at each time step within the red contour in Fig.~\ref{fig:a} to present its variation over time. The left panel shows that the wavelength-integrated Stokes $V$ signal from the \ion{Fe}{I} line reaches a maximum at about 9:40:48 UT followed by a declining phase, while the signal from the \ion{Ca}{II} line does not show a similar trend. The decrease in the \ion{Fe}{I} Stokes $V$ signal could be the result of magnetic flux cancellation due to the interaction of the opposite polarities. Now, the LP signal variation (right panel) from the \ion{Ca}{II} line stays below its 3$\sigma$ value throughout the concerned period. The LP signal of the \ion{Fe}{I} line first appears above the 3$\sigma$ value $\sim$ 7 minutes later after the initiation of flux cancellation and peaks 9.4 minutes later. The decrease in Stokes $V$ followed by an increase in LP indicates the presence of an inclined field line connectivity between the interacting opposite polarity magnetic features (see the second row of Fig.~\ref{fig:b} as well). Such an increase in LP signal during small-scale magnetic flux cancellation events occurred in the neighborhood of an evolving active region are reported in \cite{2019A&A...622A.200K}. The temporal behavior of decreasing $V$ while LP increase is consistent with the idea of the magnetic field of opposite $B_z$ undergoing reconnection and becoming more horizontal as reconnection takes place. Since the interaction is happening between an emerging magnetic feature and an existing one (see Section \ref{subsec:roi}) of opposite polarity, we suggest that the reconnection is occurring via the formation of a magnetic loop, although its precise shape cannot be determined from our observations (see Sect.~\ref{sec:cut}).\\

\begin{figure*}
\centering
  \includegraphics[trim=10 335 260 55,clip,width=1.0\textwidth]{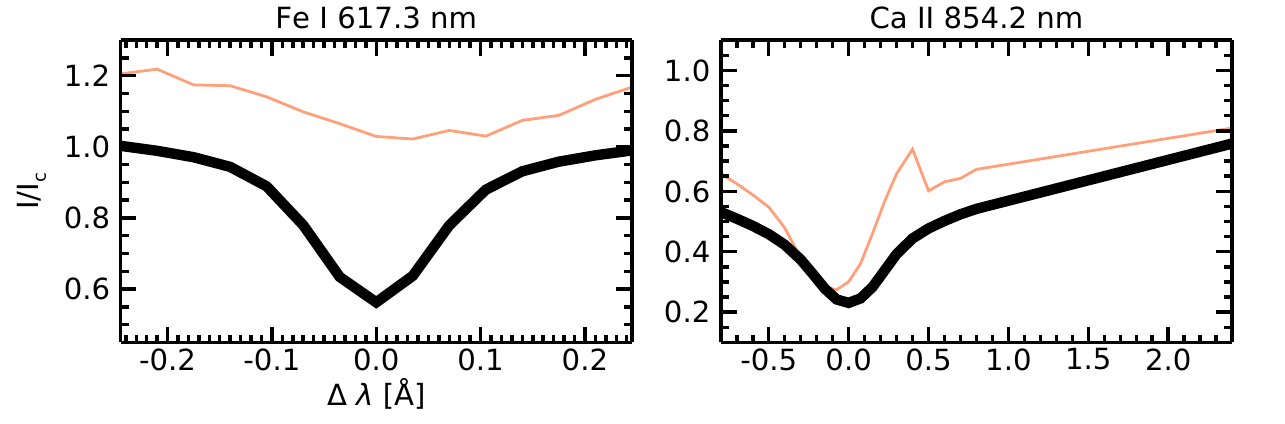}
  \caption{Left: in red is the Stokes $I$ profile of \ion{Fe}{I} at 9:44:9 UT. Right: \ion{Ca}{II} Stokes $I$ profile at 9:40:14 UT taken from the same pixel from which the profile on the left panel is made. In both cases, the profiles in black represent the spatiotemporal mean QS profile for the respective spectral lines.}
  \label{fig:e}
\end{figure*}

We chose a five-pixel region -- the pixel which shows the highest intensity on the blue wing of H$\alpha$ and four pixels around it -- to check the spectral profiles of the three observed lines during the four peaks indicated in Fig.~\ref{fig:b} (fifth row). The signature of QSEBs is coded into these profiles as shown in Fig.~\ref{fig:d} from the top (region 1) to the bottom panels (region 4). Each of the orange curves represents the spectral profiles of each of the five pixels in the selected region. The left-most column in Fig.~\ref{fig:d} shows the variation in the Stokes $I$ profiles of \ion{Fe}{I} line in comparison with the mean QS profile (solid black line). When compared to the corresponding QS signal, the continuum and core intensities of the \ion{Fe}{I} line increase by more than 20~\%, and 70~\%, respectively, during cancellation. And at times, the core of the \ion{Fe}{I} line (left panel in Fig.~\ref{fig:e}) even surpasses the QS continuum. This implies that the temperature where the core of Fe I is formed (say i.e. $\log\tau_c \approx -2$) is as high as the temperature of the quiet Sun at $\log\tau_c=0$ (i.e. $\approx 6400$~Kelvin). We observe shifts in the \ion{Fe}{I} line profiles as well -- blue shifted at 9:50 UT (first column on the fourth row of Fig.~\ref{fig:d}). To the best of our knowledge, this is the first report of a QSEB event's imprint in the \ion{Fe}{I} 6173~{\AA} line. Even though three of the data sets used in \cite{2016A&A...592A.100R} had similar spectral sampling with almost the same wavelength range for the \ion{Fe}{I} 6173~{\AA} line, they reported the absence of any QSEB-related signal in this neutral iron line.\\

The middle column in Fig.~\ref{fig:d} presents the evolution of \ion{Ca}{II} Stokes $I$ profiles at the four selected times. The wing intensity at 2.4~{\AA} rises by more than 40~\% above the QS value during the cancellation. Enhancements in a similar range are found by \cite{2018MNRAS.479.3274S} as well. Meanwhile, the line core signal of \ion{Ca}{II} becomes about 60~\% higher than the QS core intensity. Another aspect we noticed in a few Stokes $I$ profiles of the \ion{Ca}{II} line is the emission-like humps on the red wing (see right panel of Fig.~\ref{fig:e}). The signal of the strongest hump at this time is about 60~-~70~\% higher than the QS value at the corresponding wavelength point. We also report that the \ion{Ca}{II} Stokes $I$ profiles are shifted at times when compared to the QS profile; for e.g., it shows a blueshift at 9:43:45 UT (top-middle column in Fig.~\ref{fig:d}). Such shifts and core enhancement in the \ion{Ca}{II} line are also reported in the work by \cite{2018MNRAS.479.3274S}.\\

The rightmost column in Fig.~\ref{fig:d} displays the H$\alpha$ intensity profiles. As presented by \cite{2017ApJ...845...16N} our observations also show the H$\alpha$ spectral lines having enhanced inner and outer wings with an unaffected core. We also point out that the blue-wing of this spectral line has a higher intensity than the red-wing at all times. The H$\alpha$ blue wing intensity increases by almost 40~\% compared to the QS value during the flux cancellation. This is higher than the 20~\% enhancement reported by \cite{2018MNRAS.479.3274S}.\\

\begin{figure*}
  $\begin{array}{rl}
   {\includegraphics[trim=20 115 110 75,clip,width=0.5\textwidth]{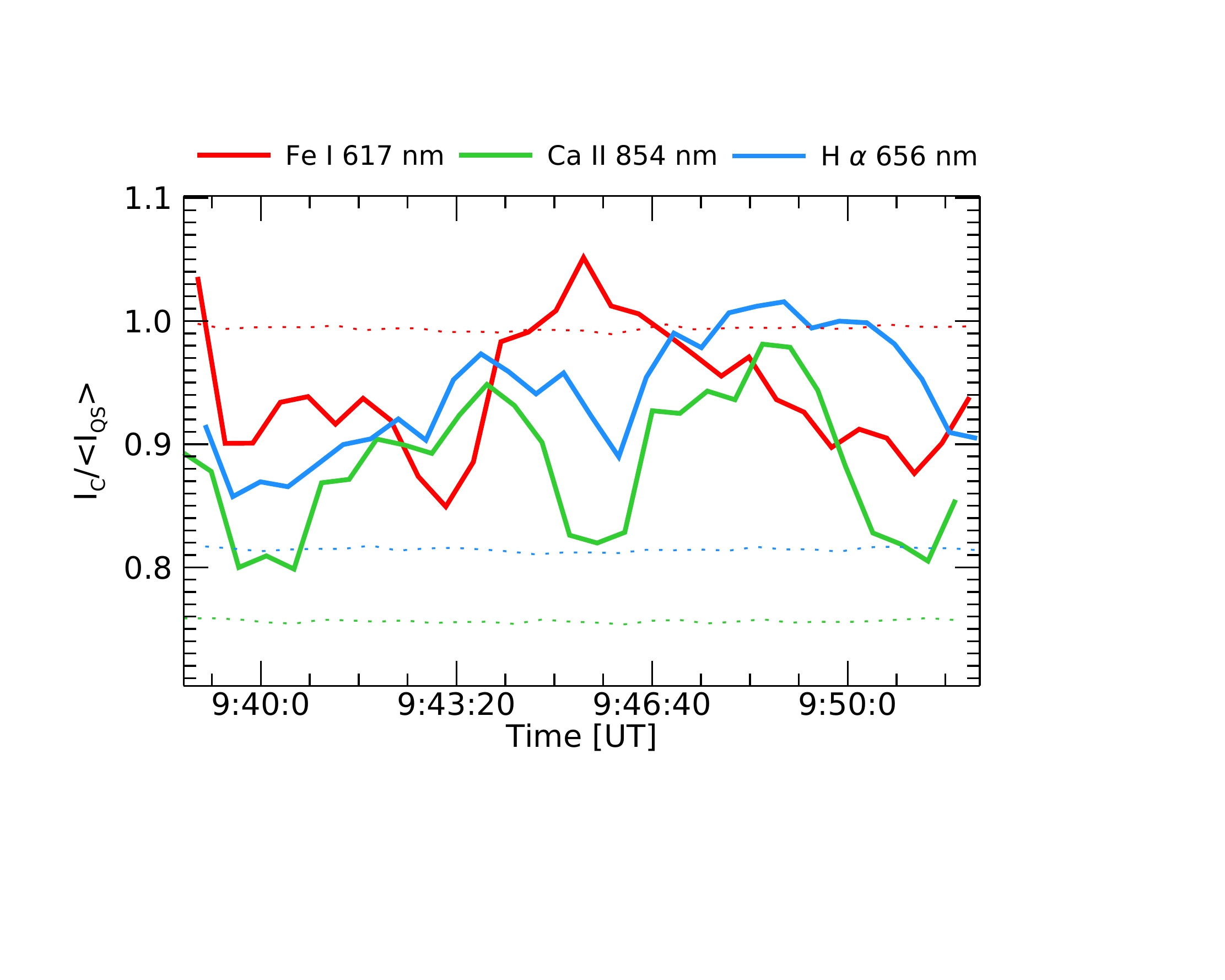}}&
   {\includegraphics[trim=20 115 110 70,clip,width=0.5\textwidth]{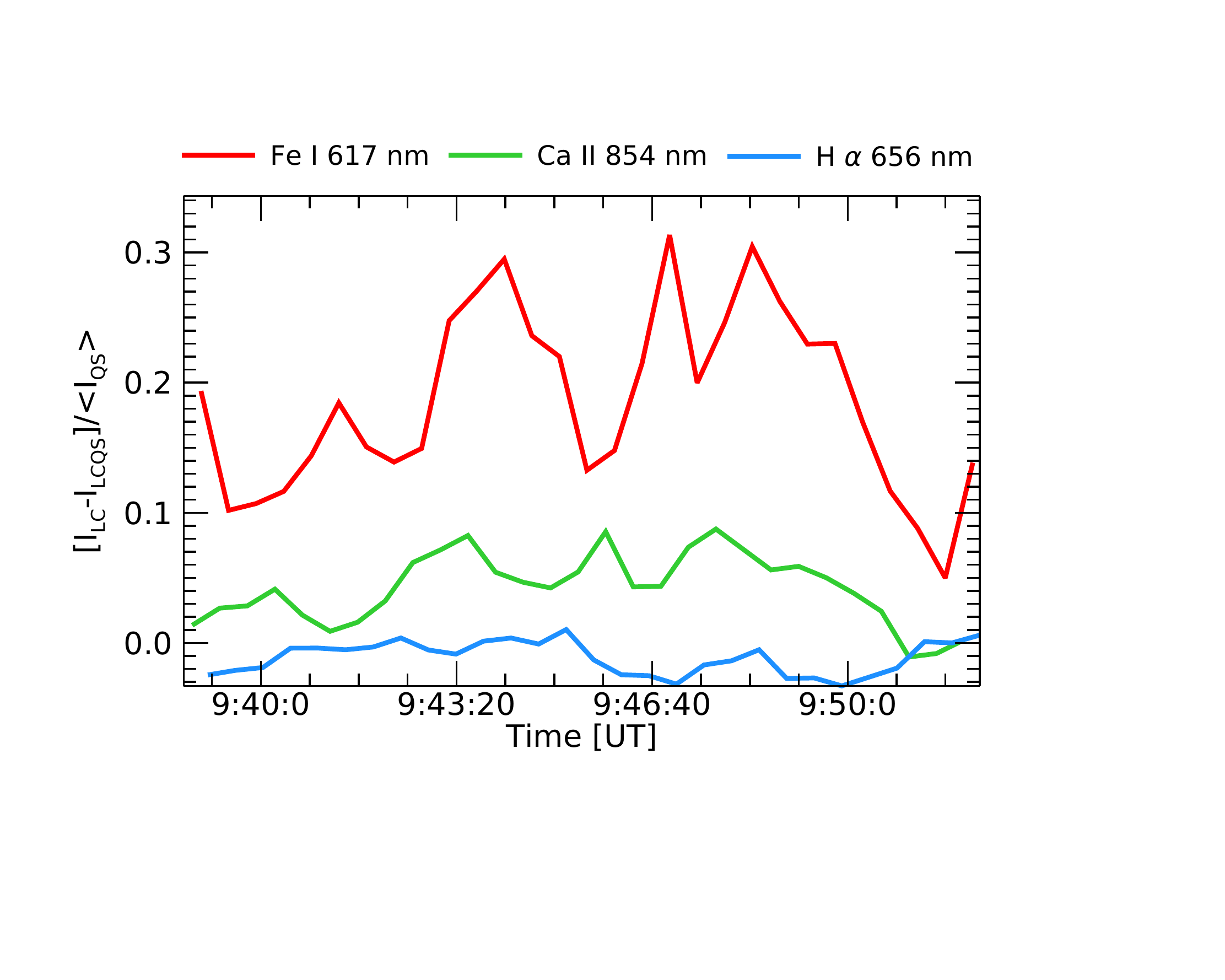}}
\end{array}$
  \caption{Plots in red, green and blue represent values for \ion{Fe}{I}, \ion{Ca}{II} and H$\alpha$, respectively. Left: continuum/wing intensity curves; Right: line core intensity evolution relative to their respective QS value. The dotted lines in the left panel represent mean quiet Sun values for the respective spectral lines.
  \label{fig:f}}
\end{figure*}

The continuum intensity of \ion{Fe}{I}, and wing intensity of H$\alpha$ and \ion{Ca}{II} averaged over the aforementioned five-pixel region is shown in the left panel plot of Fig.~\ref{fig:f}. All three lines display multiple peaks. And, the prominent peak of \ion{Fe}{I} line leads that of \ion{Ca}{II} and H$\alpha$. \cite{2018MNRAS.479.3274S} also reported repetitive enhancements in the wings of H$\alpha$ and \ion{Ca}{II} (albeit at different wavelengths from ours). In their case, the \ion{Ca}{II} peaks lead that of H$\alpha$ in most cases. While their data set had a cadence of 8 seconds, our cadence is much slower (about 28 sec; see Sect.~\ref{sec:obsv}) and therefore we are unable to draw a similar conclusion.\\

The right panel of Fig.~\ref{fig:f} shows the line core intensity evolution of the \ion{Fe}{I}, \ion{Ca}{II}, and H$\alpha$ lines. To improve the visibility of the peaks we plotted the line core intensity relative to its average QS value as a function of time. Like continuum/wing intensity plots, the line core intensity also exhibits repetitive intensity enhancements.\\

Connecting the information provided by the plots in Fig.~\ref{fig:c} and Fig.~\ref{fig:f}, we can state qualitatively that the event under consideration is giving rise to multiple instances of local heating driven by reconnection-related magnetic-flux cancellation. Section \ref{sec:invr} makes use of the results from the FIRTEZ-dz inversion runs to substantiate this proposition and quantify the physical parameters associated with the event under consideration.

\subsection{Quantitative Results}
\label{sec:invr}

\begin{figure*}
\centering
  \includegraphics[trim=15 350 105 65,clip,width=1.0\textwidth]{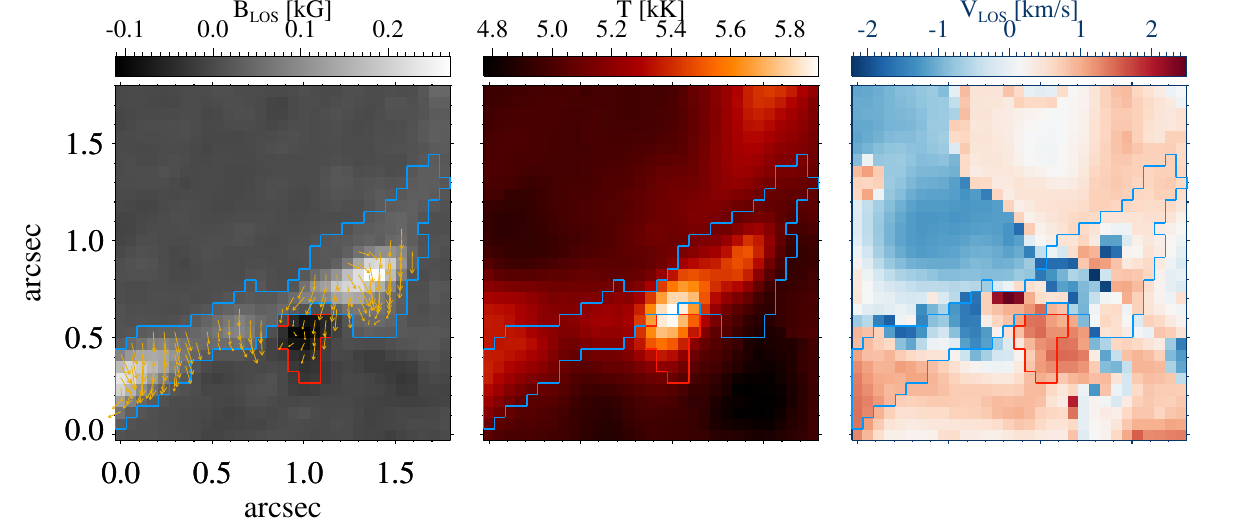}
  \caption{Left to right: maps at $\log\tau_c$~=~-1.5 of LOS component of magnetic field vector with the transverse magnetic field component over-plotted as arrows; temperature, and LOS Doppler velocity. The red and blue contours outline the negative and positive polarity patches, respectively.}
  \label{fig:g}
\end{figure*}

Fig.~\ref{fig:g} shows maps of line-of-sight (LOS) magnetic field (left), temperature (middle), and LOS Doppler velocity at $\log\tau_c \approx -1.5$ for the same FOV as shown in Fig.~\ref{fig:b}. The $B_{LOS}$ map includes an arrow field representing the ambiguity-corrected (Sect.~\ref{subsec:180amb}) transverse field component. These arrows clearly show that the magnetic field lines are indeed oriented at a tilted angle and that both polarities are magnetically connected. This is in agreement with the presence of LP signal between the opposite polarity magnetic features as shown in the second row of Fig.~\ref{fig:b}. Now, the middle panel reveals that the temperature of the region, where we observe a QSEB, is also higher than the immediate surroundings. This region encompasses both polarities as well. From the Doppler velocity map (rightmost panel in Fig.~\ref{fig:g}) we see that the region covered by the QSEB has both upflow and downflow.\\

\begin{figure*}
$\begin{array}{rl}
\centering
  {\includegraphics[trim=20 115 110 70,clip,width=0.5\textwidth]{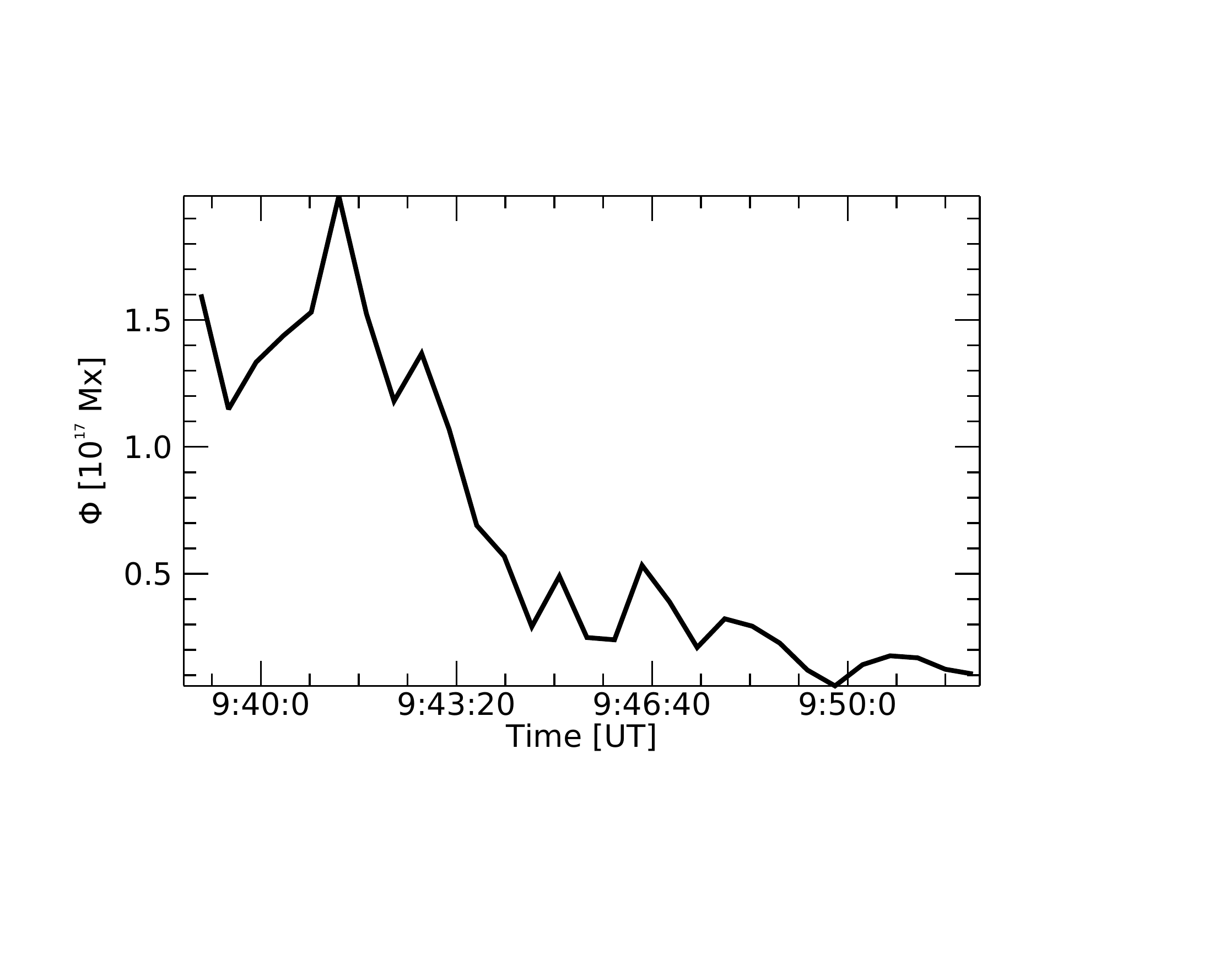}}&
  {\includegraphics[trim=16 130 15 75,clip,width=0.44\textwidth]{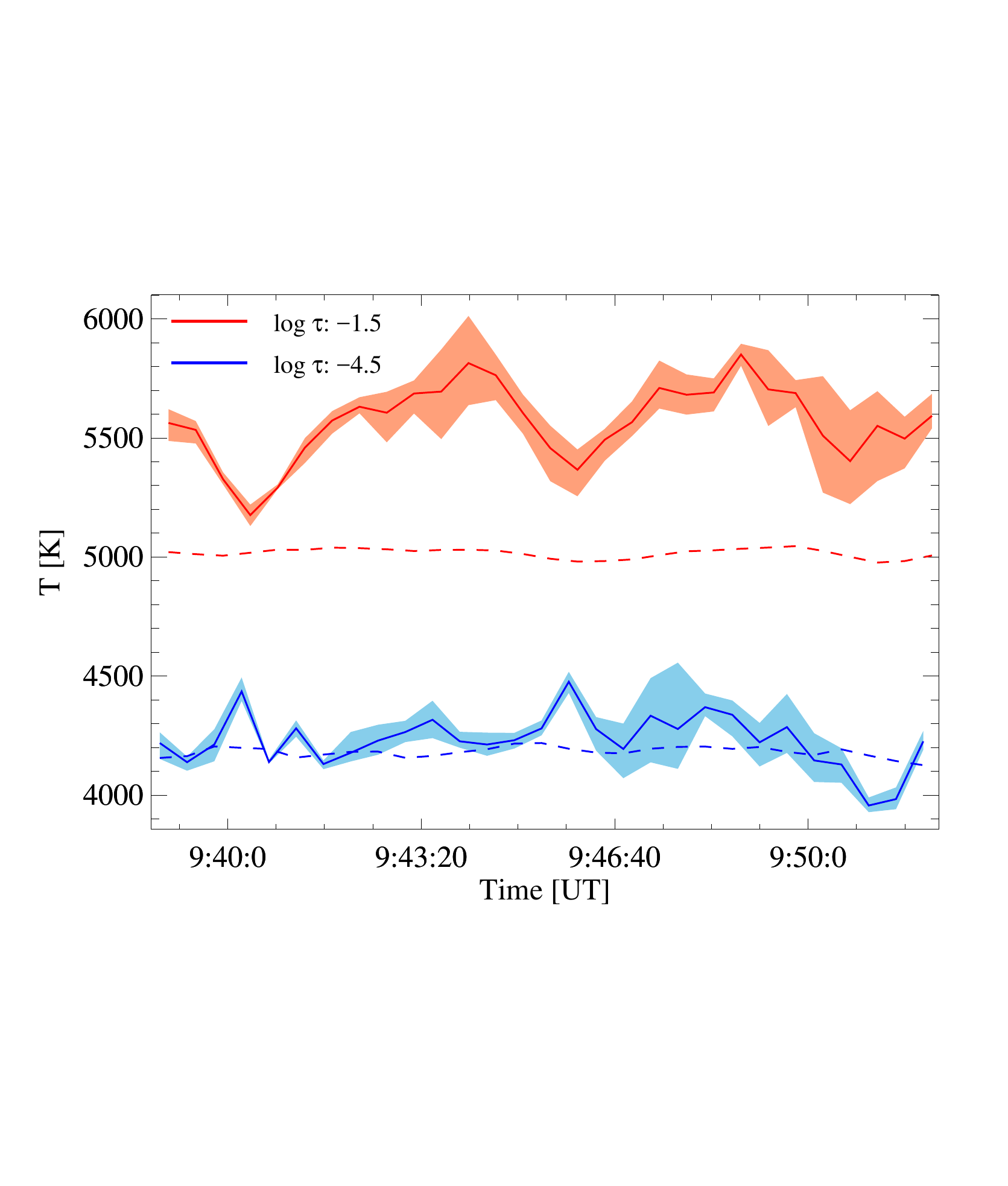}}
\end{array}$
  \caption{Left: temporal variation of vertical magnetic flux of the negative polarity patch at $\log\tau_c$~=~-1.5. We summed the magnetic flux of the pixels within the red contour shown in Fig.~\ref{fig:b} to obtain flux at each time frame. Right: Variation of temperature values for the 5 chosen pixels at $\log\tau_c$~=~-1.5 (red shaded region) and at $\log\tau_c$~=~-4.5 (blue shaded region). The solid red and blue lines represent the mean over the 5-pixel region at respective $\log\tau_c$ levels. The dashed lines in red and blue represent the mean QS temperatures at the respective $\log\tau$ values.
  \label{fig:h}}
\end{figure*}

The change in the magnetic flux of the negative patch with time is presented in the left panel of Fig.~\ref{fig:h}. The magnetic flux peaks at 9:41:20 UT and from there more than 90~$\%$ of the negative polarity flux deplete in less than 11 minutes. The flux cancellation rate is calculated to be $\sim$~${6 \times 10^{14}}$ Mx/s. \cite{2016ApJ...823..110R} have reported a flux cancellation rate in a similar range for EBs. Most of the cancellation rate is due to the area of the negative polarity patch decreasing over time, with the average $B_{LOS}$ changing a little over time. We also determined the total magnetic energy in the height range between $\log\tau_c$~=~-1.5 and $\log\tau_c$~=~-4.5 using the equation: 

\begin{equation}
E_\mathrm{B}(t)=\frac{1}{8\pi}~\oiint\limits_{A} \left\{\int_{Z_1}^{Z_2} B^2(x,y,z,t)~{\rm d}z\right\}~{\rm d}x {\rm d}y, 
\end{equation}

where $A$ is the area of the negative patch at instant t which ranges from $[10^{14} - 2.0 \times 10^{15}]$~cm$^{2}$. $Z_1$ and $Z_2$ are the geometrical heights corresponding to $\log\tau_c$~=~-1.5 and $\log\tau_c$~=~-4.5, respectively. For the considered height range ($\Delta z$~=~$<Z_\mathrm{\log\tau_c = -4.5}~-~Z_\mathrm{\log\tau_c = -1.5}>~\sim$~430 km) $E_\mathrm{B}$ of the negative patch varies between $[4\times 10^{24} - 2\times 10^{26}]$~erg. We found a systematic decline in the magnetic energy (not shown here) that presents an exactly similar trend shown by the magnetic flux of the negative patch (see left panel of Fig.~\ref{fig:h}). $E_\mathrm{B}$ peaks at 9:41:20 UT and loses 99.6\% of that value in about 11 minutes. We repeated the calculation for the heights covered between $\log\tau_c$~=~0 and $\log\tau_c$~=~-1.5 ($<Z_\mathrm{\log\tau_c~=~-1.5}~-~Z_\mathrm{\log\tau_c~=~0.0}>~\sim$~230~km) and obtained magnetic energy of the negative patch in the range of $[2\times 10^{24}~-~10^{26}]$~erg. For a height range, $\Delta z$, of $\sim$ 400 km, \cite{2016ApJ...823..110R} obtained magnetic energies in the range of $[10^{24}~-~10^{25}]$~erg. This is an order of magnitude lower than any of our estimations for $E_\mathrm{B}$.

The evolution of temperature values at $\log\tau_c$~=~-1.5 and at $\log\tau_c$~=~-4.5 for the five-pixel region centered on the peak H$\alpha$ blue wing intensity is shown in the right panel of Fig.~\ref{fig:h}. The curves present multiple peaks at both optical depths. The prominent peak in temperature at $\log\tau_c$~=~-1.5 is $\sim 1000$~K higher than the QS temperature at that instant (at 9:44:9 UT). The highest peak at $\log\tau_c$~=~-4.5 (at 9:47:45 UT) is $\sim 360$~K above the respective QS temperature.  It is important to state here that in this paper we are mostly dealing with temperature differences, and therefore, the systematic underestimation of the real temperatures under the FDC approximation (see Sect.~\ref{subsec:average_qs}), cancels out to first order of approximation in our calculations. Finally, we derived the total thermal energy needed for heating in the height range between $\log\tau_c$~=~-1.5 and $\log\tau_c$~=~-4.5 using the following equation

\begin{equation}
E_\mathrm{t}(t)=\frac{3 k_\mathrm{B}}{2} \oiint\limits_{A} \left\{\int_{Z_1}^{Z_2} n(x,y,z,t)[T(x,y,z,t)-T_{\rm QS}(z,t)]~{\rm d}z\right\}~{\rm d}x {\rm d}y,
\end{equation}

where $k_\mathrm{B}$ is the Boltzmann constant, n(x,y,z) is the number density, and [T(x,y,z,t)~-~T$_{\rm QS}$(z,t)] is the temperature enhancement at given (x,y,z,t) with respect to the mean QS temperature at given $z$ and at time $t$. From the inversion returned value of density $\rho$(x,y,z,t), we calculated n(x,y,z,t) as:

\begin{equation}
     n(x,y,z,t)=\frac{\rho(x,y,z,t)}{[1.66\times10^{-24}\times(0.91+0.09\times4)]},
\end{equation}

where $1.66\times10^{-24}$ g is the atomic mass unit. We note that we considered a mixture of only hydrogen (91\%) and helium (9\%) and have taken into account that helium is 4 times heavier than hydrogen. In the calculation of n(x,y,z,t), the number of electrons is neglected as they are about two orders of magnitude smaller in the photosphere and lower chromosphere where the plasma is only partially ionized. We obtained $E_\mathrm{t}$ in the range of $[5\times 10^{23}$~-~$4\times 10^{25}]$~erg. Comparing $E_\mathrm{b}$ and $E_\mathrm{t}$, it appears that the magnetic energy is sufficient to account for the heating in the range of heights we considered for the evaluation.

\subsection{Temporal evolution of the physical parameters in vertical slices}
\label{sec:cut}

To gain a deeper insight into the QSEB analyzed in this work, and to present its connection to the magnetic reconnection, we study the physical parameters as a function of time and height ($z$) along the first three slices indicated in Fig.~\ref{fig:b} (bottom panel). Our observations show that the detected QSEBs are located on a magnetic element that itself is on an intergranular lane. Magnetic elements are known to possess enhanced temperatures, compared to the quiet Sun, in the mid-photosphere \citep{solanki1986network,solanki1992network,lagg2010imax}. Moreover, intergranular lanes harbor large downflow velocities in the deep-photosphere that become upflows in the upper-photosphere \citep[see Fig.~3 in][]{borrero2002}. We confirmed these features by calculating the average temperature $T$$_{\rm 0}$(z), and average line-of-sight velocity $v$$_{\rm LOS,0}$(z) stratifications over regions where $\|B$$_{\rm LOS}$($\log\tau_c$~=~-1.5)$\| \ge$ $50$~Gauss in Fig~\ref{fig:b} (second row).

\begin{figure*}
\centering
  {\includegraphics[trim=2 116  75 138,clip,width=0.85\textwidth]{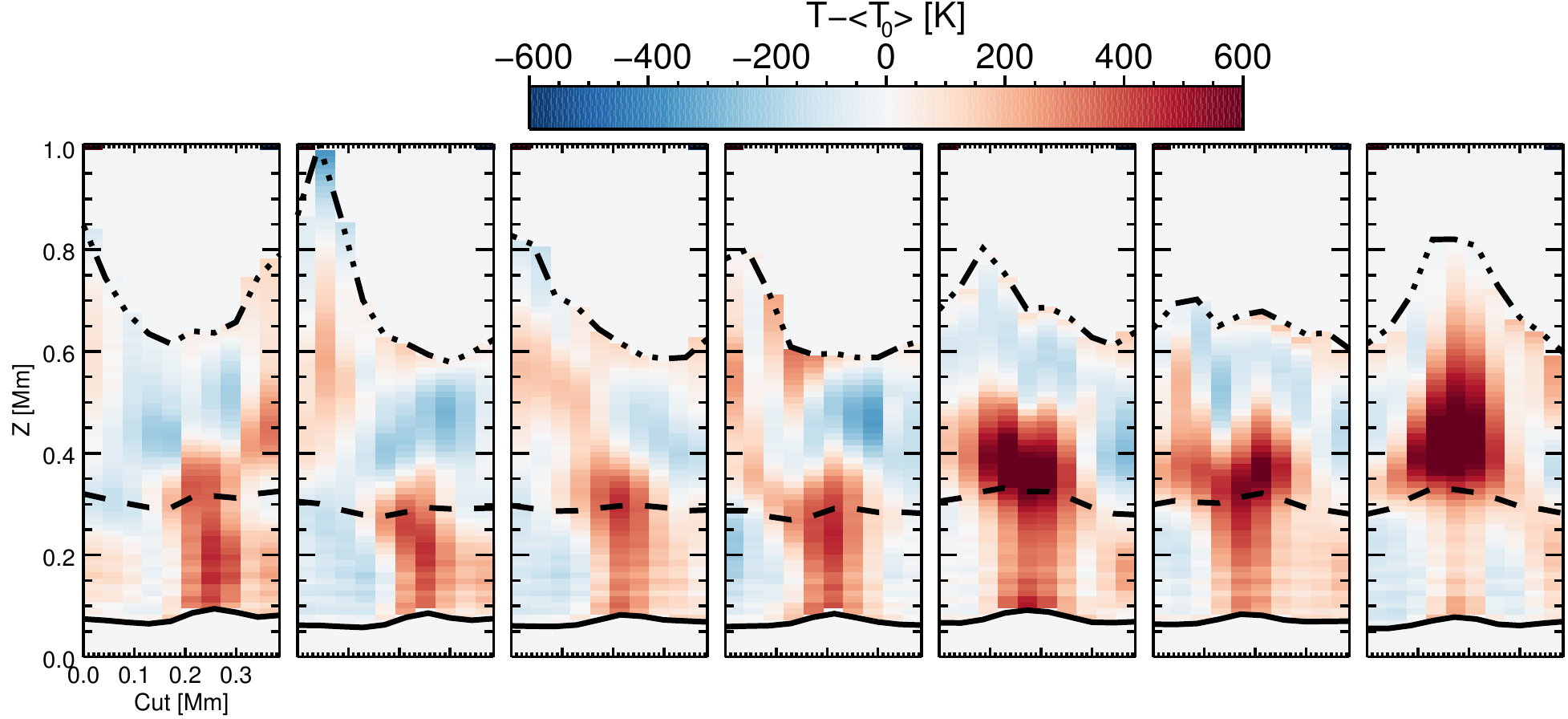}}\\
  {\includegraphics[trim=2 116  75 140,clip,width=0.85\textwidth]{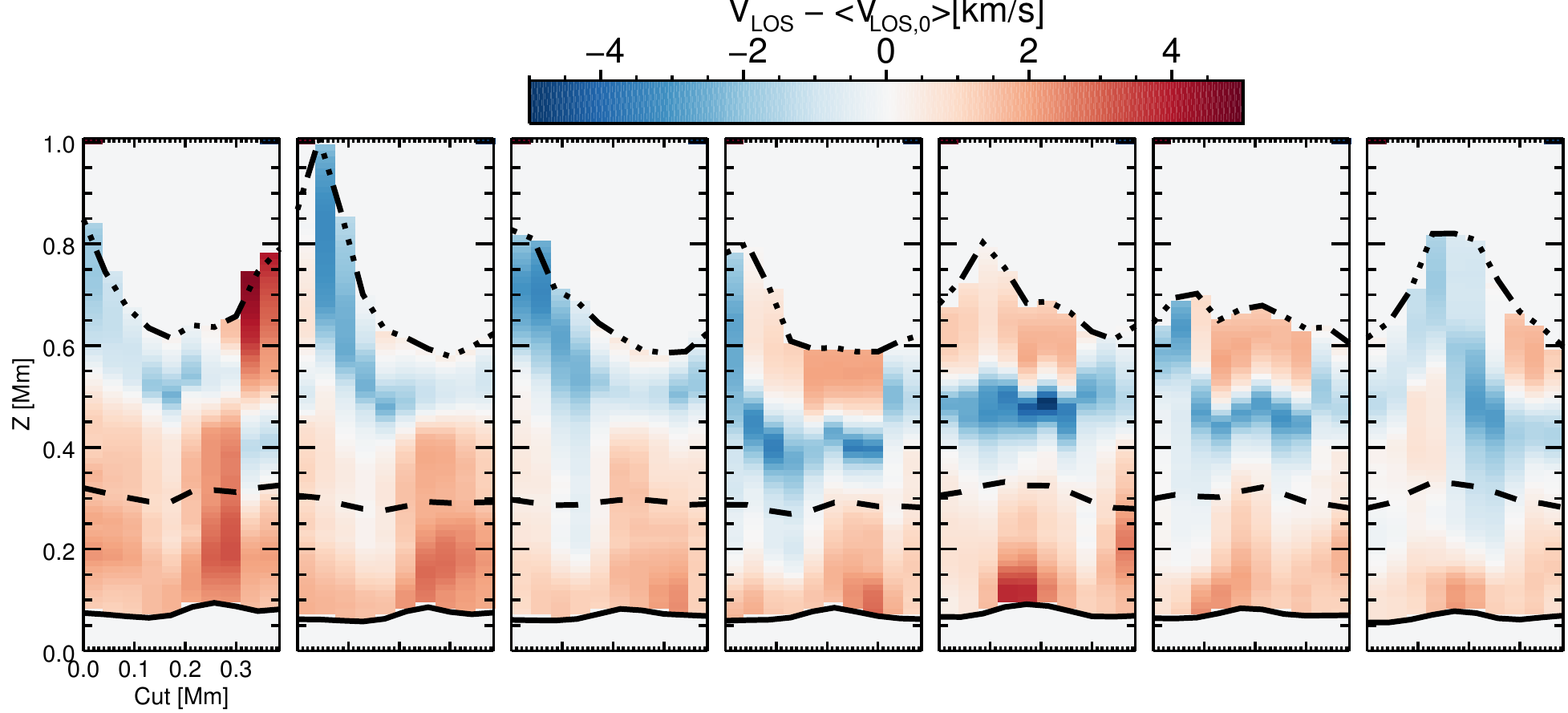}}\\
  {\includegraphics[trim=2 120  75 140,clip,width=0.85\textwidth]{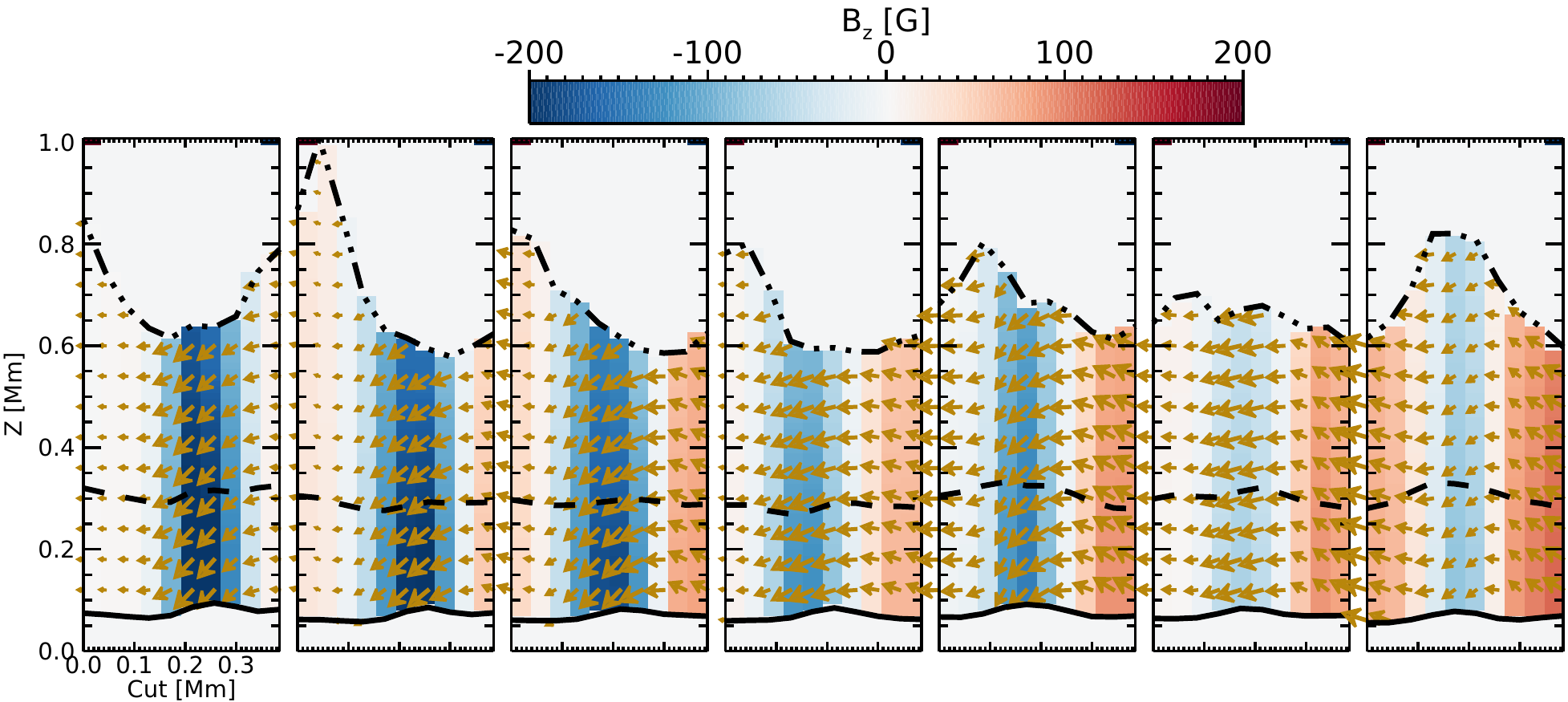}}
  \caption{{\it From top to bottom}: Evolution of temperature, Doppler velocity, and $B_\mathrm{z}$ with overlaid arrows representing the transverse magnetic field. In all the rows, the fourth map represents the $Z$ stratification of the respective physical parameter along the cut shown next to H$\alpha$ blue-wing intensity enhancement denoted as 1 in Fig.~\ref{fig:b}; zero on the X-axis represents the lower-left point of the cut. The maps before and after the fourth one are along the same cut but for time steps behind and ahead of that of the fourth one. Time runs from 9:42:16 - 9:45:05 UT (left to right). Solid, dashed, and dash-dotted black lines represent $\log\tau_c = 0, -1.5,$ and $-4.5$ lines, respectively.
  \label{fig:i}}
\end{figure*}

\begin{figure*}
\centering
 {\includegraphics[trim=2 116 75 138,clip,width=0.85\textwidth]{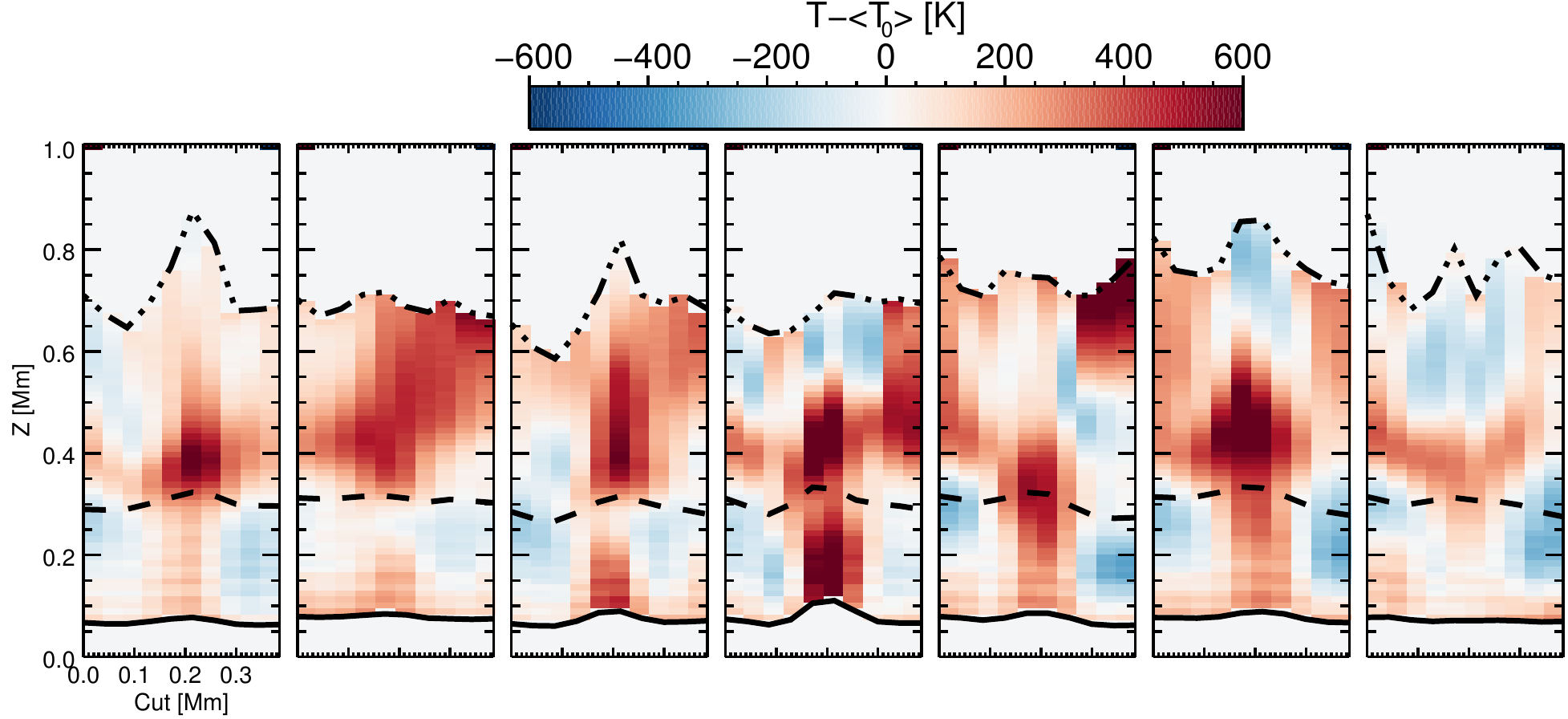}}\\
  {\includegraphics[trim=2 116  75 140,clip,width=0.85\textwidth]{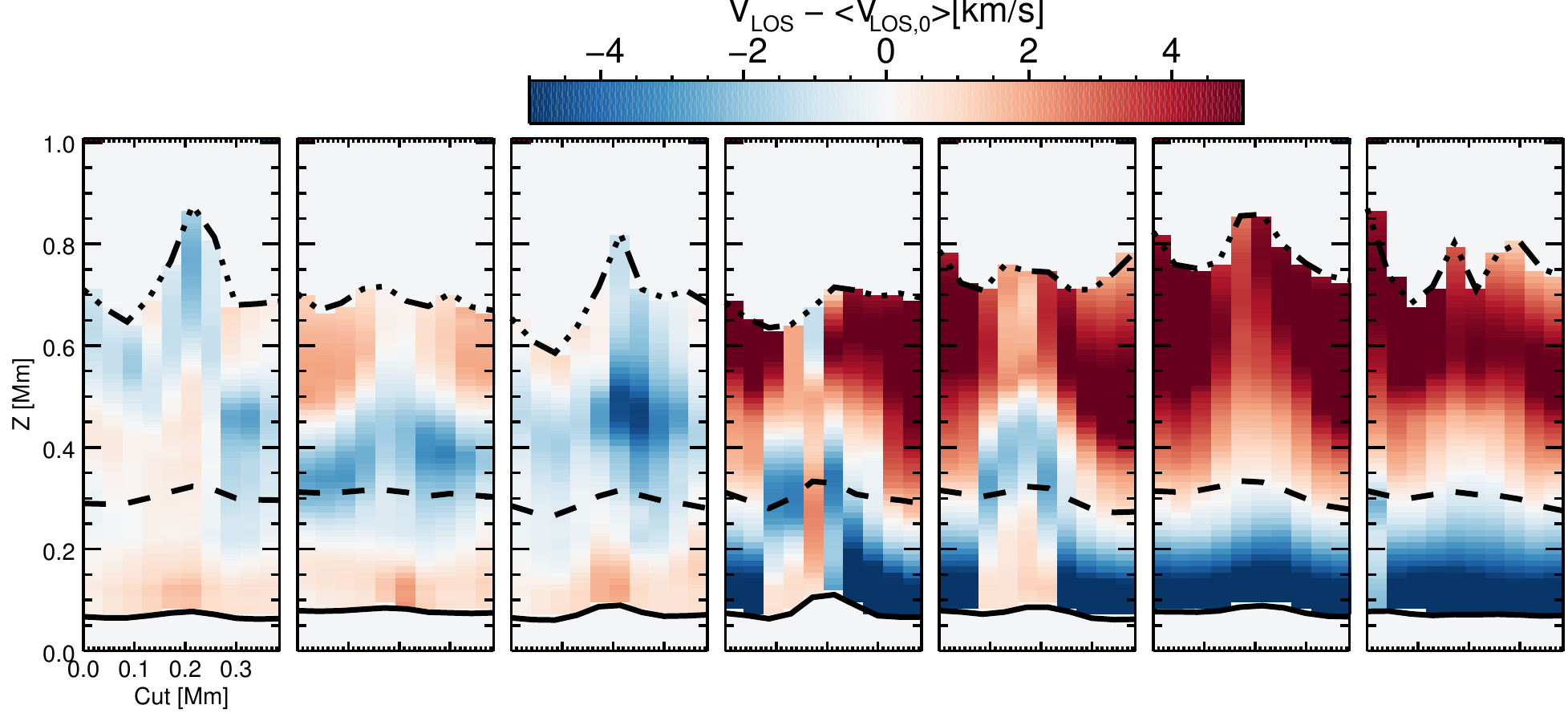}}\\
  {\includegraphics[trim=2 120  75 140,clip,width=0.85\textwidth]{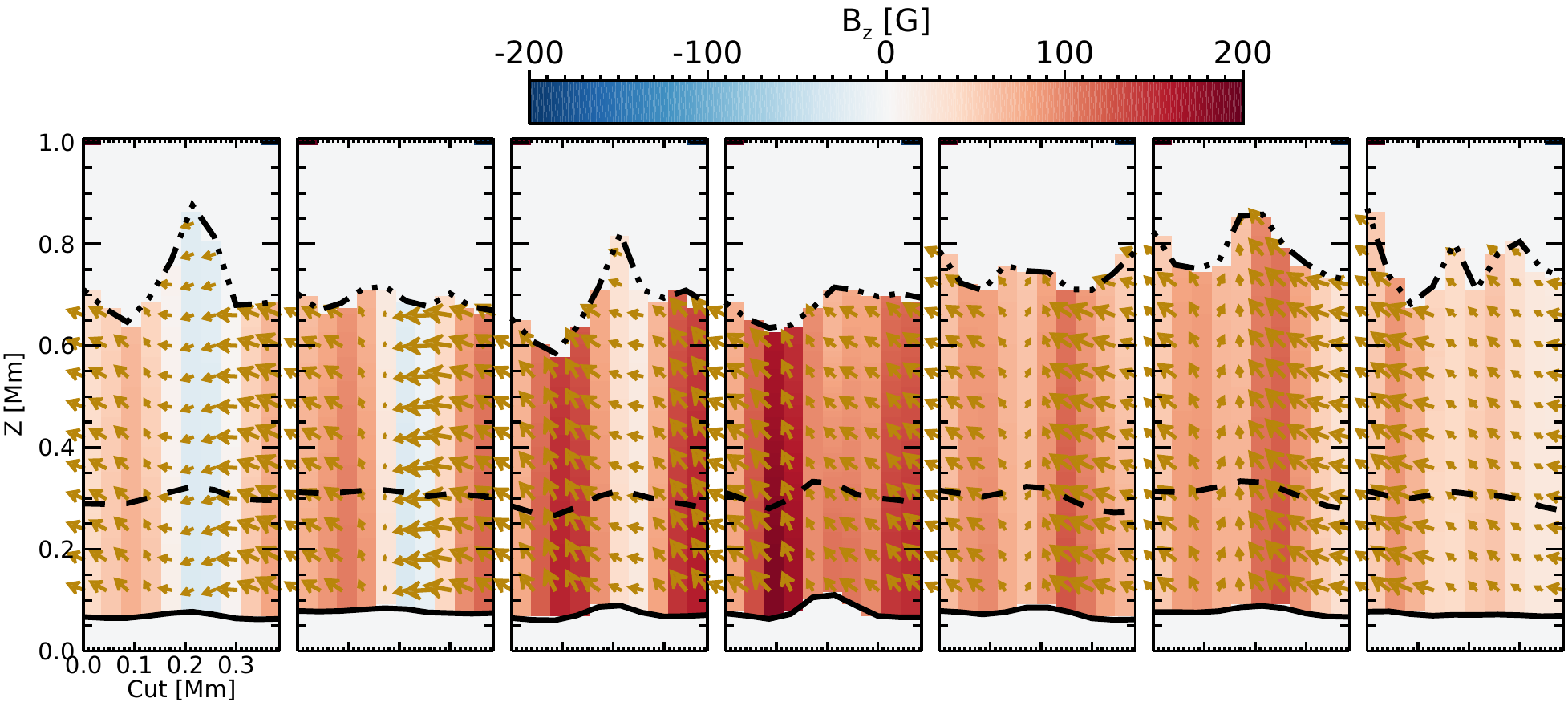}}
  \caption{Same as Fig.~\ref{fig:i} but for the cut denoted as 2 (third column, final panel) in Fig.~\ref{fig:b}. Time is from 9:45:33 - 9:48:22 UT (left to right). Solid, dashed, and dash-dotted black lines represent $\log\tau_c$=0, -1.5, and -4.5 lines, respectively.
  \label{fig:j}}
\end{figure*}

\begin{figure*}
\centering
  {\includegraphics[trim=2 116 75 138,clip,width=0.85\textwidth]{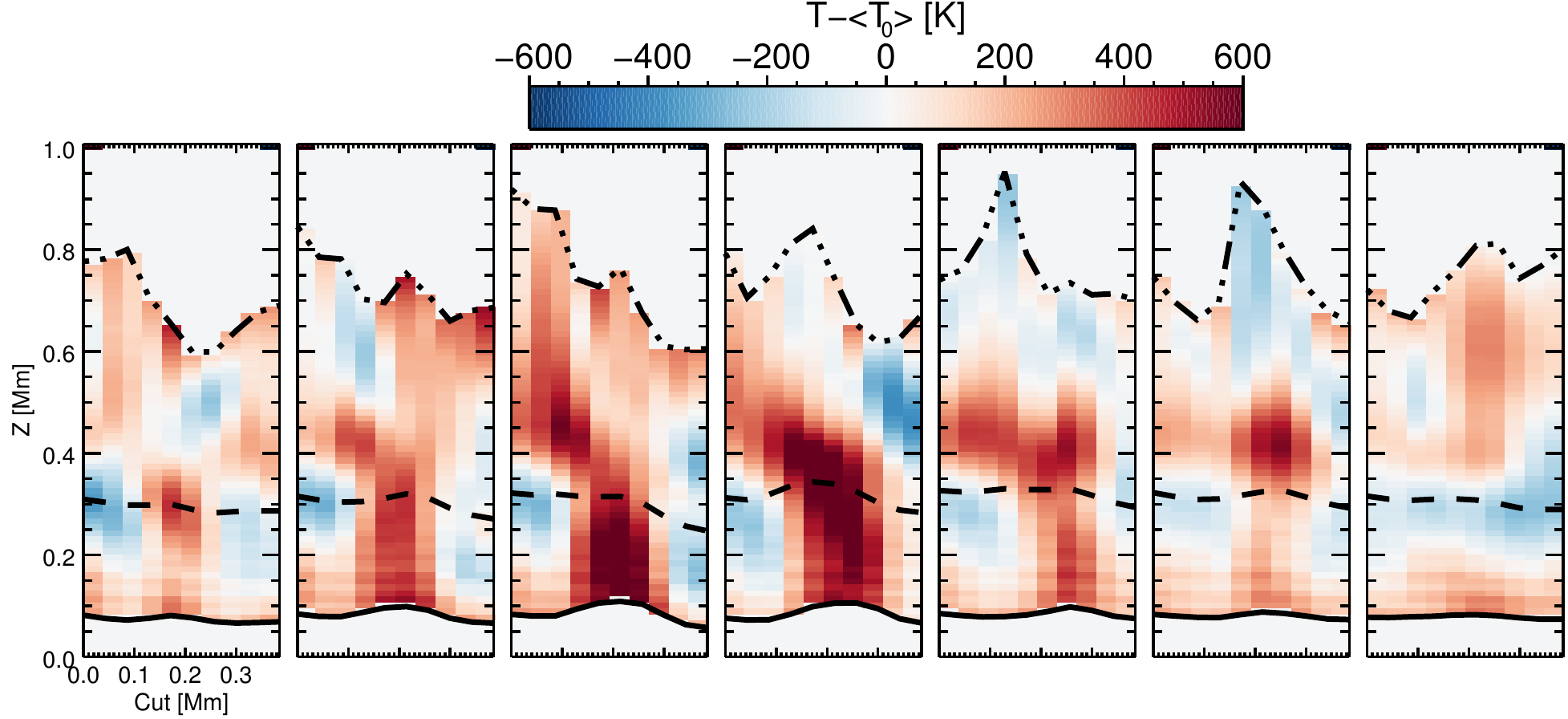}} \\
  {\includegraphics[trim=2 116 75 140,clip,width=0.85\textwidth]{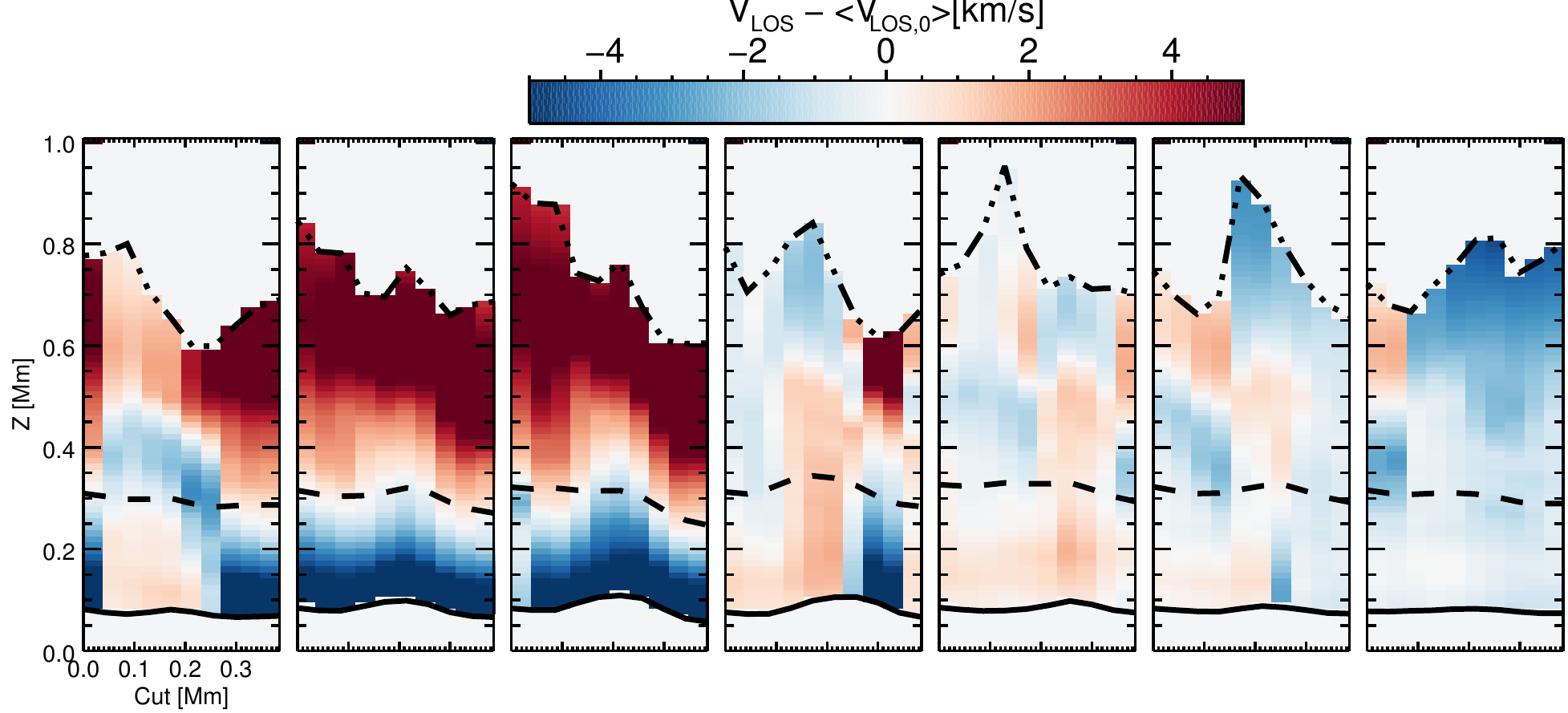}}\\
  {\includegraphics[trim=2 120 75 140,clip,width=0.85\textwidth]{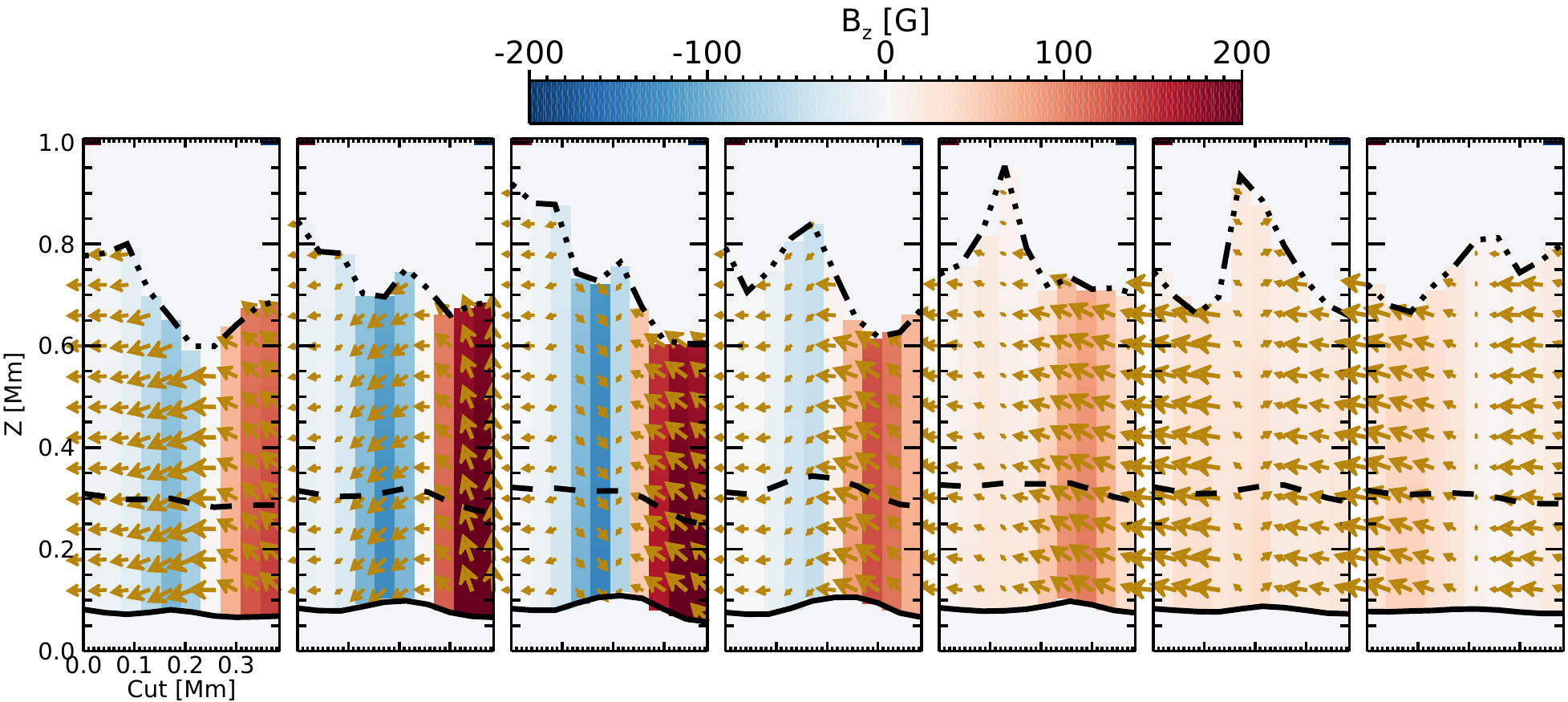}}
  \caption{Same as Fig.~\ref{fig:i}, but for the third cut shown in the fourth column of the final row in Fig.~\ref{fig:b}. In all the maps, solid, dashed, and dash-dotted black lines represent $\log\tau_c$= 0, -1.5, and -4.5 lines, respectively.
  \label{fig:k}}
\end{figure*}

Similar patterns of temperature enhancement and downflow/upflow in the deep/upper photosphere are also seen in the average stratification over the red contour in Fig.~\ref{fig:b} where the QSEBs appear. Therefore, in order to isolate the features that uniquely belong to the QSEB phenomenon from those typically occurring in an isolated magnetic element, we plot in Fig.~\ref{fig:i} (cut \#1), Fig.~\ref{fig:j} (cut \#2), and Fig.~\ref{fig:k} (cut \#3) the temperature and line-of-sight velocity along the aforementioned slices in Fig.~\ref{fig:b} (bottom panel) but relative to the average, at each height, of the isolated element: $T$(L,z) - $T$$_{\rm 0}$(z) and $v$$_{\rm LOS}$(L,z) - $v$$_{\rm 0}$(z). Here, $L$ refers to the direction of the slices in Fig.~\ref{fig:b}. The magnetic field (bottom panels in Figs.~\ref{fig:i} through ~\ref{fig:k}) is not displayed with respect to the isolated magnetic element because individual magnetic features may possess very different values of the magnetic field and therefore an average is not particularly representative. The arrow field in the bottom panels of these figures represents the projection of the magnetic field vector onto the $(L,z)$-plane. All panels display the location of the $\log\tau_c$=0, -1.5, and -4.5 in solid, dashed, and dashed-dotted lines, respectively. Regions where $\log\tau_c > 0$ and $\log\tau_c<-4.5$ are not shown because the spectral lines do not provide reliable information in those layers.

According to Fig.~\ref{fig:i} (upper panel) for cut \#1 the QSEB event leads to enhanced temperatures, with respect to the isolated element, sometimes as high as 1000~Kelvin. These temperature enhancements appear initially below the $\log\tau_c$~=~-1.5 level, but they move towards the upper-photosphere and low-chromosphere as time progresses. The location of the temperature increase is consistent with the intensity enhancements seen in the continuum/wing/line-core intensity of the different spectral lines (see Fig.~\ref{fig:b}). Upflows as large as $v_{\rm los} \approx -4$ km~s$^{-1}$ and downflows of $\approx 2$ km~s$^{-1}$ are seen above and below, respectively, the $\log\tau_c$~=~-1.5 level. The magnetic field in the $(L,z)$-plane appears continuous: i.e. the arrow field does not present abrupt interruptions.\\

Cut \# 2 is displayed in Fig.~\ref{fig:j}. Here, increased temperatures are mostly located above $\log\tau_c$~=~-1.5 (top panel). Initially, the line-of-sight velocity (middle panel) displays a similar behavior as in cut \# 1, with upflows/downflows above/below the $\log\tau_c$~=~-1.5-level. However, at some point, the velocity pattern drastically shifts: downflows turn into upflows and vice-versa, while the overall magnitude of the flows increases: both upflow and downflow are now clearly larger than 5 km~s$^{-1}$. This change is accompanied by a sudden change in the connectivity of the magnetic field lines in the bottom panel: on the fifth frame, the component of the magnetic field along the $L$-direction presents an abrupt jump around $L \approx 0.18$~Mm with $B_L$ of opposite signs facing each other.  Along this thin vertical strip where field line disconnection happens, we observe reduced Doppler velocity compared to the nearby locations. In addition, a reduction in temperature in comparison to the previous frame also occurs.\\

Cut \#3, displayed in Fig.~\ref{fig:k}, is similar to cut \#2 in that the velocity pattern of extreme downflows above $\log\tau_c$~=~-1.5 and upflows below this level are present, albeit from the start. A sudden change in the connectivity of the field lines in the $(L,z)$-plane is also seen on the third frame, again with opposite $B_L$ facing each other at around $L \approx 0.2$~Mm.\\

The velocity pattern of upflows/downflows above/below the $\log\tau_c$~=~-1.5 level (see all panels on the second row of Fig.~\ref{fig:i}) is consistent with the bi-directional jets that appear during and after reconnection \citep{pontin2022}. The inverse pattern of downflows/upflows above/below the $\log\tau_c$~=~-1.5 level (see fourth through seventh columns on the middle row of Fig.~\ref{fig:j}) does not fit, at first glance, with the idea of diverging jets caused by magnetic reconnection. However, we cannot rule out (because our inversions are not reliable there) that reconnection happens, at some point during the time evolution above the lower-chromosphere (i.e. $\log\tau_c <-4.5$) leading to the formation of strong downflows between $\log\tau_c \in [-1.5,-4.5]$. In fact, there are some hints that this might actually be happening in cut \# 2 (Fig.~\ref{fig:j}) as there seem to be strong temperature enhancements very close to, but right below, $\log\tau_c \approx -4.5$.\\

From the observed temperature enhancements, we surmise that there are multiple instances of magnetic reconnection happening at different heights, either simultaneously or at different times, and that the velocity maps display complex patterns indicating that the diverging jets occurring as a result of those reconnection events superpose each other. The changes in the flow velocity appear to be correlated with a sudden change in the connectivity of the field lines. In case magnetic field connectivity is not affected the flow pattern also remains unchanged.\\

 Concerning the geometry of the magnetic field it is not possible, with our current observations, to establish whether the QSEB presented in this work is produced by magnetic reconnection in a U \citep{2017ApJ...839...22H, 2019A&A...626A..33H} or $\Omega$-loop \citep{2017A&A...601A.122D}. This is due to the very small linear polarization signals in the Ca II line (see Fig.~\ref{fig:c}) that make it impossible to determine whether the horizontal fields are located mostly in the high or deep photosphere.

\section{Summary} \label{sec:conc}

In this work, we have analyzed a reconnection-related cancellation event and an associated quiet Sun Ellerman bomb (QSEB) observed with the 1-m Solar Swedish Telescope. The QSEB is initially detected above a negative polarity region but is observed later on the nearby positive polarity patch as well. The event presents many features that had already been previously reported \citep{2018MNRAS.479.3274S} such as enhanced intensity in the wing of \ion{Ca}{II} 854 nm line and of H$\alpha$, but enhanced line-core intensity only in the former spectral line. In addition, we find signatures related to the quiet Sun Ellerman bomb in the continuum and line-core intensity of the \ion{Fe}{I} line at 617 nm. Despite EBs and QSEBs being widely regarded as occurring in the Photosphere \citep{2011ApJ...736...71W,2016A&A...592A.100R} this is the first time its signatures are seen close to the photospheric continuum at optical wavelengths \citep[i.e. $\log\tau_c$~=~0, cf.][]{rutten2016eb}. In fact, our observations clearly show that both continuum and line-core intensity in \ion{Fe}{I} at 617 nm dramatically increase during the event, with the line-core intensity reaching higher values than the quiet Sun continuum intensity in this spectral line (see left panel Fig.~\ref{fig:e}). This can only happen if the temperature during the QSEB  at $\log\tau_c$~=~-1.5 (approximate height of formation of the line-core) is larger than the temperature on the quiet Sun at $\log\tau_c$~=~0: $T$$_{EB}$($\log\tau_c$~=~-1.5) > $T$$_{QS}$($\log\tau_c$~=~0) $\approx$ 6400~Kelvin. This does not mean that the \ion{Fe}{I} line at 617 nm turns into emission though, as its continuum intensity also greatly increases during the QSEB event.

We have further analyzed the time-series of the spectropolarimetric data during this event in order to infer the physical parameters, such as temperature, line-of-sight velocity, magnetic field, etc. as a function of $(x,y,z)$, and time $t$, by employing the FIRTEZ-dz inversion code \citep{adur2019firtez,borrero2021mhs} using non-LTE radiative transfer. The temperature shows recurrent enhancements of up to 1000~Kelvin with respect to an unperturbed magnetic element. The enhancements are seen at a variety of heights, anywhere from the deep photosphere ($\log\tau_c \approx 0$) to the lower chromosphere ($\log\tau_c \approx -4.5$). Enhanced temperatures, as well as enhanced intensities in the \ion{Fe}{I}, \ion{Ca}{II} and H$\alpha$ lines, occur several times during the event (Figs.~\ref{fig:i},\ref{fig:j},\ref{fig:k} and \ref{fig:b}), thus indicating several episodic temperature rises within the lifetime of the event.\\

The line-of-sight velocity shows the common pattern of bi-directional jets commonly ascribed to magnetic reconnection \citep{pontin2022}: large upflows in the mid-upper photosphere and lower chromosphere and large downflows in the deep photosphere. This has already been reported in Ellerman bombs \citep{2011ApJ...736...71W}. We have inferred, however, that the velocity pattern dramatically changes and the strength of the flows increases whenever
the connectivity of the field lines undergoes a sudden disconnection. These results provide the strongest evidence so far for magnetic reconnection as being the cause of quiet Sun Ellerman bombs, just as is the case for EBs \citep{2013ApJ...779..125N}.\\

Magnetic reconnection as the cause of Ellerman bombs nearby active regions had already been proposed and explored, among others, by \citet{2016ApJ...823..110R}. To explore this possibility in quiet Sun Ellerman bombs we have studied the amount of thermal energy required to heat up the atmosphere from the temperatures in the quiet Sun to the inferred temperatures and compared it to the available magnetic energy that can be potentially converted into heat via Joule dissipation. Our energy estimations show that the magnetic energy released during the flux cancellation can account for heating in the range of heights considered in this investigation, thus supporting magnetic reconnection as a potential driver for QSEBs. 

\section*{Acknowledgements}

The Swedish 1-m Solar Telescope is operated on the island of La Palma by the Institute for Solar Physics of Stockholm University in the Spanish Observatorio del Roque de los Muchachos of the Instituto de Astrof\'{\i}sica de Canarias. The Institute for Solar Physics is supported by a grant for research infrastructures of national importance from the Swedish Research Council (registration number 2017-00625). This project has received funding from the European Research Council (ERC) under the European Union’s Horizon 2020 research and innovation program (SUNMAG, grant agreement 759548). AJK thanks Philip Lindner for collecting the data as part of a backup proposal under the Trans-National Access Programme of SOLARNET, and Oleksii Andriienko for performing the data reconstruction with SSTRED.

\section*{Data Availability}
The data underlying this article will be shared on reasonable request to the corresponding author.





\bibliographystyle{mnras}
\bibliography{mybib} 








\bsp	
\label{lastpage}
\end{document}